\documentclass[fleqn,usenatbib]{mnras}

\usepackage{newtxtext,newtxmath}

\usepackage[T1]{fontenc}

\DeclareRobustCommand{\VAN}[3]{#2}
\let\VANthebibliography\thebibliography
\def\thebibliography{\DeclareRobustCommand{\VAN}[3]{##3}\VANthebibliography}

\usepackage{graphicx}	
\usepackage{amsmath, bm}	

\newcommand{\e}{\operatorname{e}}
\newcommand{\I}{\operatorname{i}}

\newcommand{\sinc}{\operatorname{sinc}}

\newcommand{\rbar}{\operatorname{cov}}

\def\rbar{\bar \rho}

\title[Shape of the density fluctuations and the CMF]{Impact of the shape of the prestellar density fluctuations on the core mass function}

\author[P. Dumond and G. Chabrier]{
Pierre Dumond$^{1}$\thanks{E-mail: pierre.dumond@ens-lyon.fr} and Gilles Chabrier$^{1,2}$\thanks{E-mail: chabrier@ens-lyon.fr}
\\
$^{1}$CRAL, Ecole normale sup\'erieure de Lyon, Universit\'e de Lyon, UMR CNRS 5574, F-69364 Lyon Cedex 07, France\\
$^{2}$School of Physics, University of Exeter, Exeter, EX4 4QL, UK
}

\date{Accepted XXX. Received YYY; in original form ZZZ}

\pubyear{\the\year{}}

\begin{document}
\label{firstpage}
\pagerange{\pageref{firstpage}--\pageref{lastpage}}
\maketitle

\begin{abstract}
	It is well known that departure from sphericity in the geometry of primordial dark matter halos modifies their mass function. The ellipsoidal collapse model yields a better agreement with simulations of hierarchical clustering than the original, spherical model. In the present paper, we examine the same issue in the context of star formation by studying the impact of non-sphericity of density perturbations in a gravoturbulent medium on the prestellar core mass function (CMF). An important question, notably, is to find out how ellipsoidal, instead of spherical, initial density fluctuations modify both the high-mass and low-mass tails of the CMF.
	Our study shows that triaxial density fluctuations indeed depart from a purely spherical form but the deformation (prolateness and ellipticity) remains modest, suggesting that the usual hypothesis of spherical collapse in existing theories of the IMF is reasonable. We find that, as in the cosmological case, the departure from sphericity increases the collapse barrier, stabilizing the prestellar cores. The striking difference between the stellar case and the cosmological one for the ellipsoidal collapse model is that, although in both cases the less dense structures are the most deformed, they correspond to small scales, thus low mass halos in cosmology but to large scales, thus large mass cores in star formation. 
	As a result, the high mass range of the CMF is the most affected by the ellipsoidal collapse, resulting in a slightly less steep slope than the one predicted with the spherical hypothesis and a peak slightly shifted toward lower masses.  
\end{abstract}




\section{Introduction}

Despite the success of the \cite{Press_FormationGalaxiesClusters1974} formalism in deriving the mass function of primordial dark matter halos, it predicted fewer high-mass and more low-mass objects than found in numerical simulations of hierarchical clustering (e.g. \citealt{Lacey_MergerRatesHierarchical1994}). This discrepancy was significantly reduced by considering ellipsoidal rather than spherical collapse for primordial bounded structures in gaussian random fields of density fluctuations (e.g. \citealt{Sheth_EllipsoidalCollapseImproved2001, Sheth_ExcursionSetModel2002}). The physical reason for this is that, in the ellipsoidal model, the collapse barrier depends on the mass of the halos, and hence on the variance of the density field, whereas in the spherical case it does not. Such a result was expected since perturbations in a gaussian density field are inherently triaxial \citep{Doroshkevich_SpatialStructurePerturbations1973,Bardeen_StatisticsPeaksGaussian1986}, making the collapse of density perturbations more difficult. 

The spherical collapse approximation is inherently inaccurate. 
Inspired by the cosmological result, we investigate the effect of the geometry of the collapsing fluctuations in star-forming clumps on the resulting prestellar core mass function (CMF). Indeed, current theories of the CMF, which deal with the hierarchical formation of bound structures in a gravo-turbulent medium, assume that the initial density fluctuations are spherical (e.g. \cite{Hennebelle_AnalyticalTheoryInitial2008} (HC08 hereafter) and further, \citealt{Hopkins_ExcursionsetModelStructure2012} (HC12 hereafter)). 
However, it has been shown that turbulence in the ISM, in particular in the star-forming, cold, molecular phase produces highly non-spherical structures, including sheets and filaments \citep{Andre_FilamentaryNetworksDense2014, Federrath_UniversalityInterstellarFilaments2016,Arzoumanian_CharacterizingPropertiesNearby2019}. Such structures can be characterized by their fractal dimension \citep{Federrath_FRACTALDENSITYSTRUCTURE2009, Beattie_RelationTrueObserved2019}. These filaments probably shaped by turbulence \citep{Federrath_SonicScaleInterstellar2021} are believed to play a significant role in the star formation process \citep{Pineda_BubblesFilamentsCores2022,Hacar_InitialConditionsStar2022}.
It is therefore important to relax the assumption of sphericity of the perturbations and examine the implications of departing from the spherical collapse model. 

The paper is organised as follows.  
In Sect. \ref{Ellipsoid_pertub} we compute the Probability Density Function (PDF) of density fluctuations resulting from ellipsoidal perturbations induced by compressible turbulence. In particular, we study how the shape of the perturbations depends on their scale and density. In Sect. \ref{CMF_ellipsoidal_collapse} we compute the collapse condition associated with the formation of non-spherical density fluctuations. Finally, we examine the implications of this ellipsoidal collapse model on the CMF, in particular how these geometrical aspects modify the high-mass and low-mass tails of the CMF. In Sect. \ref{cosmo} we discuss the differences between the cosmological and star formation cases of the ellipsoidal collapse model. Section \ref{Discussion} is devoted to the conclusion.

We emphasize that the present paper deals only with the geometrical aspect of the formation of initial density perturbations in a star-forming cloud, induced by compressible turbulence.
In contrast to cosmological studies of halo formation (e.g. \citealt{Bond_PeakPatchPictureCosmic1996,Sheth_EllipsoidalCollapseImproved2001,Ludlow_FormationCDMHaloes2014}), we do not consider the influence of the tidal field on the collapse. This aspect has been investigated in a dedicated study (Dumond \& Chabrier 2024, subm.).

\section{Distribution of ellipsoidal perturbations in a turbulent medium}
\label{Ellipsoid_pertub}

We consider a cloud with a uniform background density within which turbulence generates density perturbations that are not assumed to be spherical, as seen in observations and numerical simulations \citep{Andre_FilamentaryNetworksDense2014, Federrath_UniversalityInterstellarFilaments2016}. The triaxial geometry of these perturbations can be modelled by ellipsoids. It is characterised by the three eigenvalues of the deformation tensor or, equivalently, the ellipticity $e$, the prolateness $p$ and the sum of the eigenvalues \(x\). In the following, we present the hypothesis on which this derivation is based and describe the computation of the joint probability of \(e, p\) and \(x\). 

\subsection{Description of the logdensity field by a stochastic random gaussian field}

The formation of density structures induced by gaussian perturbations has been extensively studied by \cite{Doroshkevich_SpatialStructurePerturbations1973} and \cite{Bardeen_StatisticsPeaksGaussian1986}. In the cosmological case, density fluctuations with respect to the background density $\rbar$ are induced by inflation and are well described by a gaussian random field. They are small ($\delta_\rho=(\rho-\rbar)/\rbar\ll1$) and can be considered as linear perturbations. 
In star formation, turbulence plays an important role, and it has been found both numerically and observationally that non-self-gravitating supersonic isothermal turbulence leads to density fluctuations that are gaussian on all scales in terms of the {\it logarithm} of the density, $s=\log(\rho/\rbar)$ \citep{Passot_DensityProbabilityDistribution1998, Kritsuk_StatisticsSupersonicIsothermal2007, Federrath_ComparingStatisticsInterstellar2010}, which precludes a perturbative approach as used in cosmology. 
From a statistical point of view, this is due to the fact that in the cosmological case the density fluctuation field is the sum of a large number of independent random additive processes. In the stellar case, it is generally accepted that the density fluctuations result from a series of multiplicative processes due to the shock-driven turbulent cascade (Raiser-Zeldovich condition). According to the central limit theorem, the first process yields a gaussian distribution, while the second yields a lognormal one for the density field \(\rho\) (e.g. \citealt{Vazquez-Semadeni_HierarchicalStructureNearly1994}).
Thus, the logarithm of the density, \(s=\ln(\rho/\rbar)\), is a stochastic random field well described by the following distribution \(\mathcal{P}_s\): 
\begin{equation}
    \mathcal{P}_s (s) = \frac{1}{\sqrt{2\pi\sigma_s^2}} \exp\left(-\frac{(s+\frac{\sigma_s^2}{2})}{2\sigma_s^2}\right),
\end{equation}
where \(\sigma_s\) is the standard deviation of the logdensity field. Assuming statistical homogeneity of the field \(s\), the PDF of \(s(\bm{q})\) is the same at any point \(q\) of the cloud.

To determine the probability distribution of the shape of the density perturbation formed in the vicinity of a local maximum in the stochastic field \(s\), we follow the same approach as in \cite{Bardeen_StatisticsPeaksGaussian1986}. To do this, we assume not only that the one-point PDF of the field \(s\) is gaussian at each point of the system, but also that \(s\) is a gaussian random field. This means that the joint n-point PDF of the random vector \((s(\bm{q}_1), ..., s(\bm{q}_n))\) is gaussian for any integer \(n\). The statistical properties of the field \(s\) are thus completely determined by its power spectrum. The relevance of this hypothesis for the description of a turbulent field is discussed in section \ref{Sec_gaussian}.

\subsection{Link between the deformation tensor and the geometry of the perturbation}
\label{Sec_link_geometry}

To study the formation of structures of different sizes and scales, often referred to as the hierarchical structure of star formation (e.g. HC08, H12), we need to study the statistics of the field \(s\) at different scales. As in HC08, we use a sharply truncated window function \(W_R(k)\) in Fourier space:
\begin{equation}
    W_R(|\bm{k}|)=\Theta(R^{-1}-|\bm{k}|),
\end{equation}
where \(\Theta\) is the Heaviside function and \(R\) is the smoothing length. This window function allows us to take into account all density fluctuations formed on scales larger than \(R\).
It also allows us to consider the correlation between the turbulence modes, since the smallest scales are completely determined by the largest ones through the turbulent cascade.  Finally, a sharp k-space filter approximation simplifies the analytical calculations. We have verified that the use of a gaussian window function \citep{Bardeen_StatisticsPeaksGaussian1986, Bond_PeakPatchPictureCosmic1996} does not alter any of the results of the present study. Based on our assumption of a gaussian random field, we deduce that the smoothed logdensity field at scale $R$ at a point \(\bm{q}\) of the cloud, defined as
\begin{equation}
    s_R(\bm{q})=s(\bm{q})\ast W_R(\bm{q})
\end{equation}
also follows a normal distribution characterised by a scale-independent mean equal to \(-\sigma_s^2/2\) and a scale-dependent variance \(\sigma_s^2(R)\), which we note as follows \(s_R(\bm{q})\hookrightarrow \mathcal{N}(-\sigma_s^2/2, \sigma_s^2(R))\). The dependence of the variance on the smoothing scale \(R\) is determined by the power spectrum of the logdensity field \(s\).

Following the same approach as in \cite{Bardeen_StatisticsPeaksGaussian1986}, and noting that \(s_R\) is also a gaussian random field based on the assumption that the local logdensity field \(s\) is gaussian, one can derive the PDF of the eigenvalues of the 3D deformation tensor, \(\mathbb{I}_{ij}^R=\partial_i\partial_j s_R\). They are well defined because \(s\) is infinitely differentiable and determine the shape of the ellipsoid. The differentiability is due to the fact that the correlation function \(g_s^R(\Delta q)\) (see eq. \ref{Eq_corr_func_app}) is infinitely differentiable in 0 (see theorem 2.2.2 in \citealt{Adler_GeometryRandomFields1981}), which is ensured by the functional form of the power spectrum characterising the statistical properties of the random field \(s\). 

We consider an overdensity formed at the scale \(R\), which reaches its maximum \(\rho_0\) at a point $\bm{q}_0$. The form of this density is assumed to be closed, i.e. the surfaces defined by $\rho_R(\bm{q})={\rm const}$ are topologically equivalent to spheres. This should be true for $\rho(\bm{q})$ relatively close to $\rho_0$. Since \(s_R\) and \(\rho_R\) are connected by an isomorphism, we can describe the shape of the perturbation with either variable. Given the homogeneity and isotropy of the background, we assume $\bm{q}_0=0$. Near the maximum we can write a Taylor expansion in basis of the three eigenvectors of $\mathbb{I}_{ij}^R$: 
\begin{equation}
    s_R(\bm{q})=s_R(0)-\frac{1}{2}\sum_{i=1}^{3}\lambda_i^R q_i^2 + O(q_i^3),
\end{equation}
where $\lambda_i^R$ are the opposite values of the associated eigenvalues. Three positive \(\lambda_i^R\) correspond to a local maximum of the density. In the following we will remove the index \(R\) for the eigenvalues, since they always depend on the smoothing scale $R$. The statistical properties of the random fields $\lambda_i$, which are linked to the statistical properties of the semi-axes of the ellipsoid, are determined in the section \ref{Shape_distribution}.

To define the size of the perturbation, we consider all the matter contained in the closed surface $\mathcal{S}=\{\bm{q}\in \mathbb{R}^3, s_R(\bm{q})=\ln(\rho_0/\rho_{\rm min})\}$, where the relative difference $\Delta \rho_0=(\rho_0-\rho_{\rm min})/\rho_{\rm min}$ is arbitrarily defined. \(\rho_{\rm min}\) is the minimum density of the perturbation. This gives the following equation:
\begin{equation}
    1=\sum_{i=1}^{3}\frac{\lambda_i}{2\ln(\rho_0/\rho_{\rm min})}q_i^2,
\end{equation}
which describes an ellipsoid, whose semi-major axes are given by : 
\begin{equation}
    \label{semi_axis}
    a_i^2=\frac{2\ln(\rho_0/\rho_{\rm min})}{\lambda_i}.
\end{equation}
By imposing \(\rho_{\rm min}\) and knowing the distribution of the eigenvalues, it is possible to determine the distribution of the geometry of the ellipsoidal fluctuations induced by the turbulent flow. In the following we will choose $\Delta \rho_0=0.1$, i.e. $\rho_0/\rho_{\rm min}=1.1$. This allows us to treat the density perturbation as a homogeneous ellipsoid, i.e. with uniform density. Given the logarithmic dependence of the closed surface \(\mathcal{S}\) on $\Delta \rho_0$, the choice of the ratio $\rho_0/\rho_{\rm min}$ has little effect on the resulting shape of the ellipsoid. We have indeed verified that the results of the calculations are almost unchanged by choosing instead \(\rho_0/\rho_{\rm min}=1.5\). In the following, we impose the following order: \(\lambda_1>\lambda_2>\lambda_3\).

\subsection{Shape distribution}
\label{Shape_distribution}

Because of the gaussian random field assumption for the logdensity \(s\), all statistical properties of this field are determined by its power spectrum.
Simulations have shown that the power spectrum \(P(k)\) of the latter is always close to \(k^{-4}\) (in 3D) (see \cite{Hennebelle_TurbulentMolecularClouds2012} \S4.2.1). Some simulations have shown that the slope of the power spectrum of the logdensity may depend on the type of forcing for the turbulence, ranging from \(k^{-3.6}\) for solenoidal driving to \(k^{-4.3}\) for compressive driving (see Appendix A in \cite{Federrath_ComparingStatisticsInterstellar2010}). Since the exact nature of the driving in the interstellar medium is not known, to choose \(k^{-4}\) between the two limits seems to be a reasonable choice.

Turbulence is assumed to be injected at the scale \(2\pi/L_{\rm i}\), with \(L_{\rm i}\) the injection length, and to dissipate at very small scale, typically smaller than the sonic scale or the ambipolar diffusion scale, as highlighted by observations \citep{Miville-Deschenes_ProbingInterstellarTurbulence2016, Pineda_ProbingPhysicsStarFormation2024}. The injection scale of turbulence is typically the size of the molecular cloud. The expression of the correlation function \(g_s^R\) of the smoothed logdensity field \(s_R(\bm{q})\) is given by : 
\begin{align}
    g^R_s(\Delta q) &= \int_{\mathbb{R}^3} P(\bm{k})W_R(|\bm{k}|) \e^{-\I\bm{k}\cdot\bm{\Delta q}} {\rm d}^3\bm{k}  \\
    &=\int_{a=\frac{2\pi}{L_{\rm i}}}^{b=\frac{2\pi}{R}}Ck^{-4}\sinc(k\Delta q)k^2{\rm d}k,
\end{align}
with \(C\) a constant that does not depend on the dissipation scale because it is much smaller than the injection scale (see Appendix \ref{App_Correlation_function}). Because we assume that the statistical field \(s\) is statistically homogeneous, the correlation length \(g^R_s\) depends only on the spatial lag \(\Delta q\). To determine the statistical properties of the semi-axes of the studied ellipsoid around a local logdensity maximum, we first need to compute the ones of the first and second derivative of the field \(s\). Both are also gaussian random fields and, as detailed in Appendix \ref{App_Correlation_function}, the dispersions of \(s_R\), \(\eta_i^R=\partial_i s_R\) and \(\mathbb{I}_{ij}^R=\partial_i\partial_j s_R\) read respectively:

\begin{align}
    \label{Sigma_0}
    \sigma_s^2(R) &= \sigma_s^2\left(1-\frac{a}{b}\right), \\
    \label{Sigma_1}
    \sigma_\eta^2(R) &= \sigma_s^2(ba-a^2), \\ 
    \label{Sigma_2}
    \sigma_\mathbb{I}^2(R) &= \frac{\sigma_s^2}{3}(ab^3-a^4),
\end{align}
with \(a=2\pi/L_{\rm i}\), \(b=2\pi/R\).
Instead of using the eigenvalues of the tensor, we define the parameters:
\begin{equation}
    \label{eq_ell_prol}
     x=\frac{(\lambda_1 + \lambda_2 + \lambda_3)}{\sigma_\mathbb{I}} \ ; \ e=\frac{\lambda_1 - \lambda_3}{2x\sigma_\mathbb{I}} \ ; \ p=\frac{\lambda_1 - 2 \lambda_2 + \lambda_3}{2x\sigma_\mathbb{I}},
\end{equation}
where \(e\) and \(p\) define respectively the ellipticity and the prolateness.  While \(e\) quantifies the 3D deformation, \(p\) gives information on the nature of this deformation: \(p>0\) corresponds to sheet like structures and \(p<0\) to filament like structures. These parameters thus give an insight of the variation of the shape of the perturbation with its density and its scale.

Following \cite{Bardeen_StatisticsPeaksGaussian1986} (appendix A and C), the joint probability distribution of the random vector \((s, x, e, p)\) imposing that \(s\) is a local density maximum (also called density peak by Bardeen) at the scale \(R\) is given by:
\begin{multline}
    \label{Joint_g_x_e_p}
    \mathcal{P}_{R}(s, x, e, p) = \frac{5^{5/2}3^{1/2}}{(2\pi)^3\sigma_s(R)}\left(\frac{\sigma_\mathbb{I}(R)}{\sigma_\eta(R)}\right)^3\frac{1}{(1-\kappa^2)^{1/2}}\e^{-Q} \\ \times x^8e(e^2-p^2)(1-2p)[(1+p)^2-9e^2]\,\chi(e,p),
\end{multline}
where
\begin{equation}
    Q =\frac{\left(s+\frac{\sigma_s^2}{2}\right)^2}{2\sigma_s^2(R)}+\frac{(x-x_\star)^2}{2(1-\kappa^2)}+\frac{5x^2}{2}(3e^2+p^2)
\end{equation}
and
\begin{equation}
    x_\star=\frac{\kappa}{\sigma_s(R)}\left(s+\frac{\sigma_s^2}{2}\right) \ ; \ \kappa=\frac{\sigma_\eta^2(R)}{\sigma_\mathbb{I}(R)\sigma_s(R)}.
\end{equation}
Here \(\chi\) is a characteristic function such that $\chi=1$ if \(3e-p<1\) and \(-e<p<e\), and $\chi=0$ otherwise. It corresponds to \(\lambda_1>\lambda_2>\lambda_3>0\), chosen to ensure the correct ordering of the semi-axes of the ellipsoid. The density probability given by equation \ref{Joint_g_x_e_p} depends only on the statistical properties of the random field \(s(\bm{q})\), namely its power spectrum, through the dependence on \(R\) of the variances \(\sigma_s^2(R), \sigma_\eta^2(R), \sigma_\mathbb{I}^2(R)\). 
It gives the statistical properties of the shape of the ellipsoid with peak density and scale. 

Looking only at the behaviour of the exponential term, we see that the lower the value of \(x\), the higher the probability that \(e\) and \(p\) be large. Assuming that \(x\simeq x_\star\), the most distorted structures are obtained when \(s(R)/\sigma_s(R)\), i.e.
the ratio of the peak density of the ellipsoid to the variance of the density fluctuations at the scale $R$, and $\kappa$ are small. We conclude that {\it the less dense structures are the most distorted}. 

\begin{figure*}
    \centering
    \includegraphics[width=\hsize]{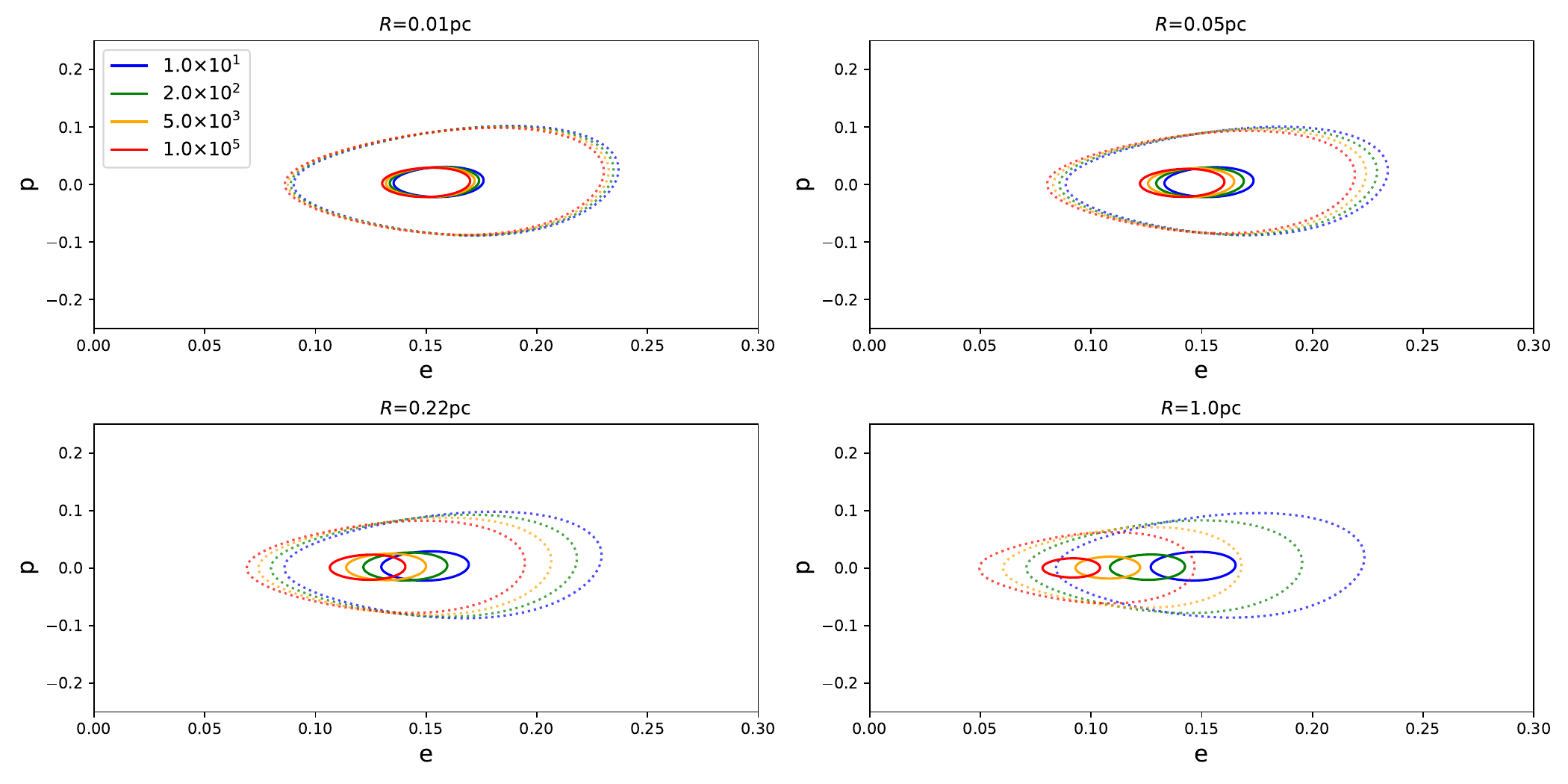}
       \caption{Density probability of the shape of the ellipsoid, \(\mathcal{P}(e, p|s)\), for several smoothing scales \(R\) and maximum densities $\rho_0/\rbar$  (indicated in the legend). The dotted and solid lines represent  the 50\% and the 95\% confidence level contours of the conditional probability, respectively. The injection length for turbulence is \(L_{\rm i}=10\text{pc}\) corresponding to a variance \(\sigma_s = 1.9\).
               }
          \label{Plot_Shape_distribution}
    \end{figure*}

These trends are confirmed by the numerical calculation of \(\mathcal{P}_R(e, p|s)\), whose density probability contour lines are plotted in Fig. \ref{Plot_Shape_distribution}. At each scale, the denser the ellipsoid, the less it is deformed. At small scales the deformation is very weakly dependent on density. For the largest, most distorted structures, the prolateness is always $<$0.1 at 50\% confidence level and only a few percent at 95\% confidence level, while the ellipticity varies from $\sim$0.05-0.1 to $\sim$0.1-0.2 for a density contrast \({\rho_0}/{\bar{\rho}}\) varying between 10 and 10$^5$. We also observe that the deformation of the ellipsoid is limited by an asymptotic value that is reached more quickly at small scales than at large ones.
We have verified that considering larger scales with larger injection lengths does not quantitatively change the result. 

Within the HC08 formalism of prestellar core density fluctuations, the less dense fluctuations correspond to the largest clumps leading to the most massive prestellar cores, while the densest clumps are the smallest. We note that, according to the shape distribution shown in Fig. \ref{Plot_Shape_distribution}, we expect the deformation of the virialised structures to be roughly independent of scale. Therefore, in contrast to the cosmological case, ellipsoidal collapse will have little effect on the mass distribution, and hence the CMF. When it does, it affects the high-mass spectrum more than the low-mass one. We examine this result in more detail in the next section.

It should be noted that the present formalism assumes gaussian statistics for the fluctuations and thus neglects higher order correlations.
Although we give arguments to justify such an approach in section \ref{Sec_gaussian}, one way to verify the accuracy of this assumption would be to perform numerical simulations to infer the statistical properties (especially ellipticity and prolateness) of the density peaks. We compare the prediction of our model to the measured distribution of ellipticity and prolateness measured in a turbulent box simulation in section \ref{Sec_Numerical_verification}.
Before this comparison, it is worth noting that our results suggest an aspect ratio \(a_3/a_1\) for the spheroidal density fluctuations on the subparsec scale of about $\sim$2. This is in very good agreement with observations of (turbulence-induced) clumps, as given in Table 1 of \cite{Tanaka_HCNHNC132020} (see also \citealt{Hacar_InitialConditionsStar2022, Krieger_TurbulentGasStructure2020}), which give aspect ratios between 1 and 3, with an average between 1.5 and 2. Our prediction is also in good agreement with the results of \cite{Lomax_IntrinsicShapesStarless2013}, who also predict an aspect ratio of the order of 2. We note that \cite{Ganguly_UnravellingStructureMagnetized2023} find that the structures formed in their turbulent zoom-in simulations are mostly sheets and filaments. However, this study essentially considers parsec-scale structures with large ratio \(\rho_0/\rho_{\rm min}\). Our formalism may not be valid any more in that case because the Taylor expansion on which it is based becomes dubious. In contrast, in the present study we consider the formation of the structures at smaller (core) scales, and we restrict our study close to the perturbation density maximum.

\subsection{Numerical verification of the model}
\label{Sec_Numerical_verification}

In this section we want to compare the distribution of ellipticity \(e\) and prolateness \(p\) given in eq. \ref{Joint_g_x_e_p} predicted by the model with the one measured in a turbulent simulation. We use the hydrodynamical code RAMSES \citep{Teyssier_CosmologicalHydrodynamicsAdaptive2002} without adaptive mesh refinement (fixed carterian grid). Boundary conditions are periodic. The numerical method is based on a second order Godunov solver scheme. The initial conditions consist of a fluid of molecular hydrogen at rest, characterised by a mean molecular weight of \(\mu=2.4\). The fluid follows an isothermal equation of state with \(T=10\)K. Because we want to test a model describing a turbulent flow without self-gravity, self-gravity is not activated in the simulation. 
The turbulence is forced using the Ornstein-Uhlenbeck forcing on the acceleration \citep{Eswaran_ExaminationForcingDirect1988a,Schmidt_NumericalDissipationBottleneck2006a,Schmidt_NumericalSimulationsCompressively2009}. The turbulence is forced at scale \(L_{\rm box}/7\). This ensures that there is enough realisation of the turbulent flow in the box to make the statistical measurements relevant. We also chose a solenoidal fraction of 0.5, which corresponds to the energy equipartition of the velocity field: 1/3 compressive and 2/3 solenoidal. The Mach number in the simulation is close to \(\mathcal{M}\simeq 5\) and the logdensity variance is \(\sigma_s\simeq 1.7\). The resolution of the simulation is 1024\(^3\). In the appendix \ref{App_Convergence} we present the results of a 512\(^3\) simulation for the convergence study. 

Once the turbulence is fully developed \citep{Federrath_UniversalitySupersonicTurbulence2013}, we extract the density structures formed with the Python package \texttt{astrodendro} \citep{Robitaille_ExposingPluralNature2019a}. We impose that each axis should be resolved by at least 8 cells and that \(\rho_{\rm max}/\rho_{\rm min}=1.5\). We chose a larger value than suggested in \ref{Sec_link_geometry} to ensure that the number of extracted structures is large enough to compute a relevant statistics of the shapes. We show examples of the extracted structures in Fig. \ref{Fig_extracted_structures}. In appendix \ref{App_Convergence}, we show that the choice of these parameters has little effect on the calculated statistics of ellipticity and prolateness.

\begin{figure*}
    \centering
    \includegraphics[width=\hsize]{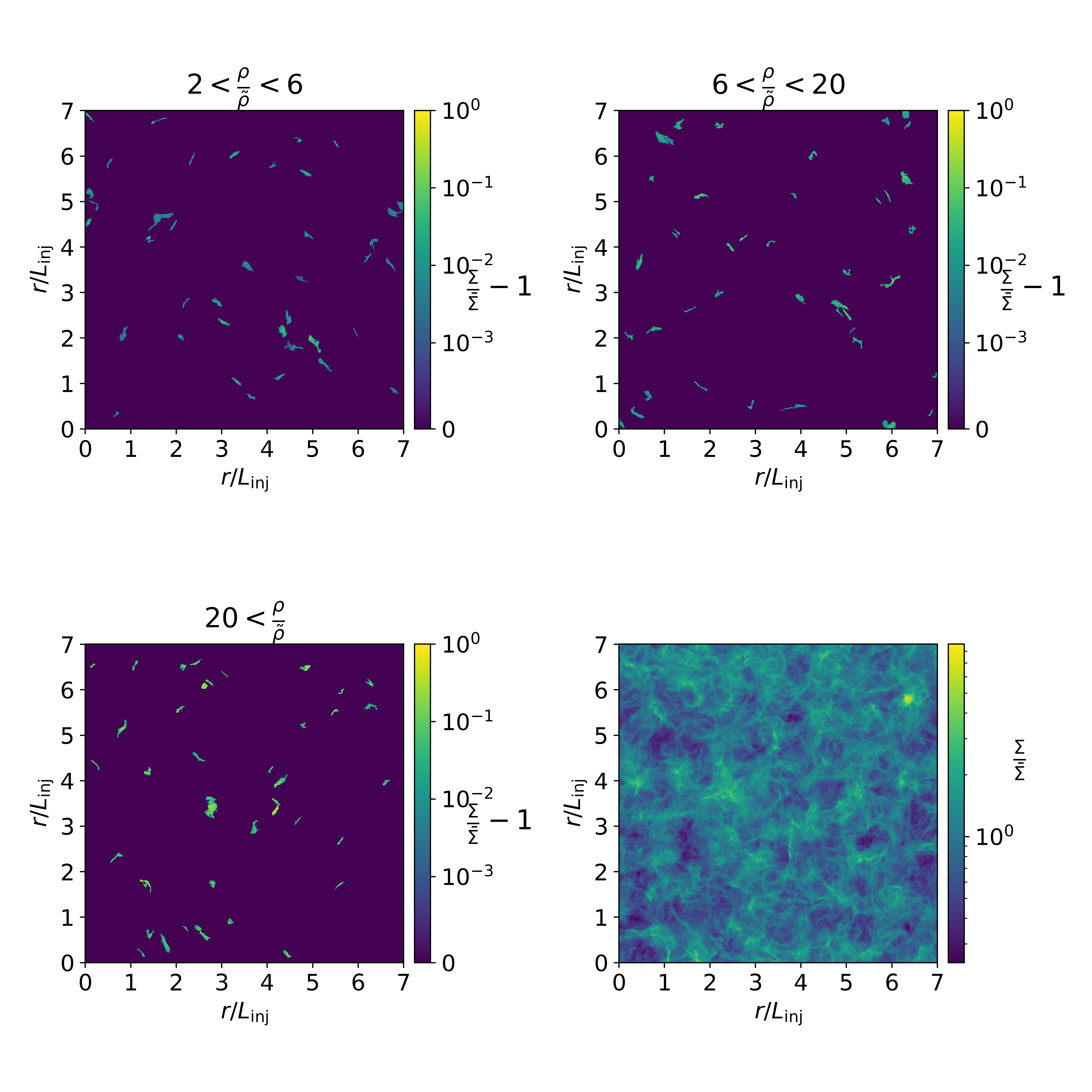}
       \caption{Example of extracted density structures for different density ranges. We plot the surface density normalized to the mean surface density \(\bar{\Sigma}\). We have replaced the value of all cells that do not belong to a structure with the average density of the simulation box to avoid polluting the line of sight to a structure. The bottom right panel is the surface density field.}
          \label{Fig_extracted_structures}
\end{figure*}

To measure the ellipticity and prolateness of the extracted structures, we first need to compute their half-axis. We follow the method presented in \cite{Ganguly_SILCCZoomDynamicBalance2024}. We first measured the moment of inertia along each axis and then calculated the three semi-axis as if the structure was an ellipsoid.
From eq. \ref{semi_axis} and \ref{eq_ell_prol}, the ellipticity and prolateness can be rewritten in terms of the semi-axis of the structure as:
\begin{equation}
    e = \frac{\frac{1}{a_1^2}-\frac{1}{a_3^2}}{2\left(\frac{1}{a_1^2}+\frac{1}{a_2^2}+\frac{1}{a_3^2}\right)} \ ; \ p = \frac{\frac{1}{a_1^2}-\frac{2}{a_3^2}+\frac{1}{a_3^2}}{2\left(\frac{1}{a_1^2}+\frac{1}{a_2^2}+\frac{1}{a_3^2}\right)} .
\end{equation}

\begin{figure*}
    \centering
    \includegraphics[width=\hsize]{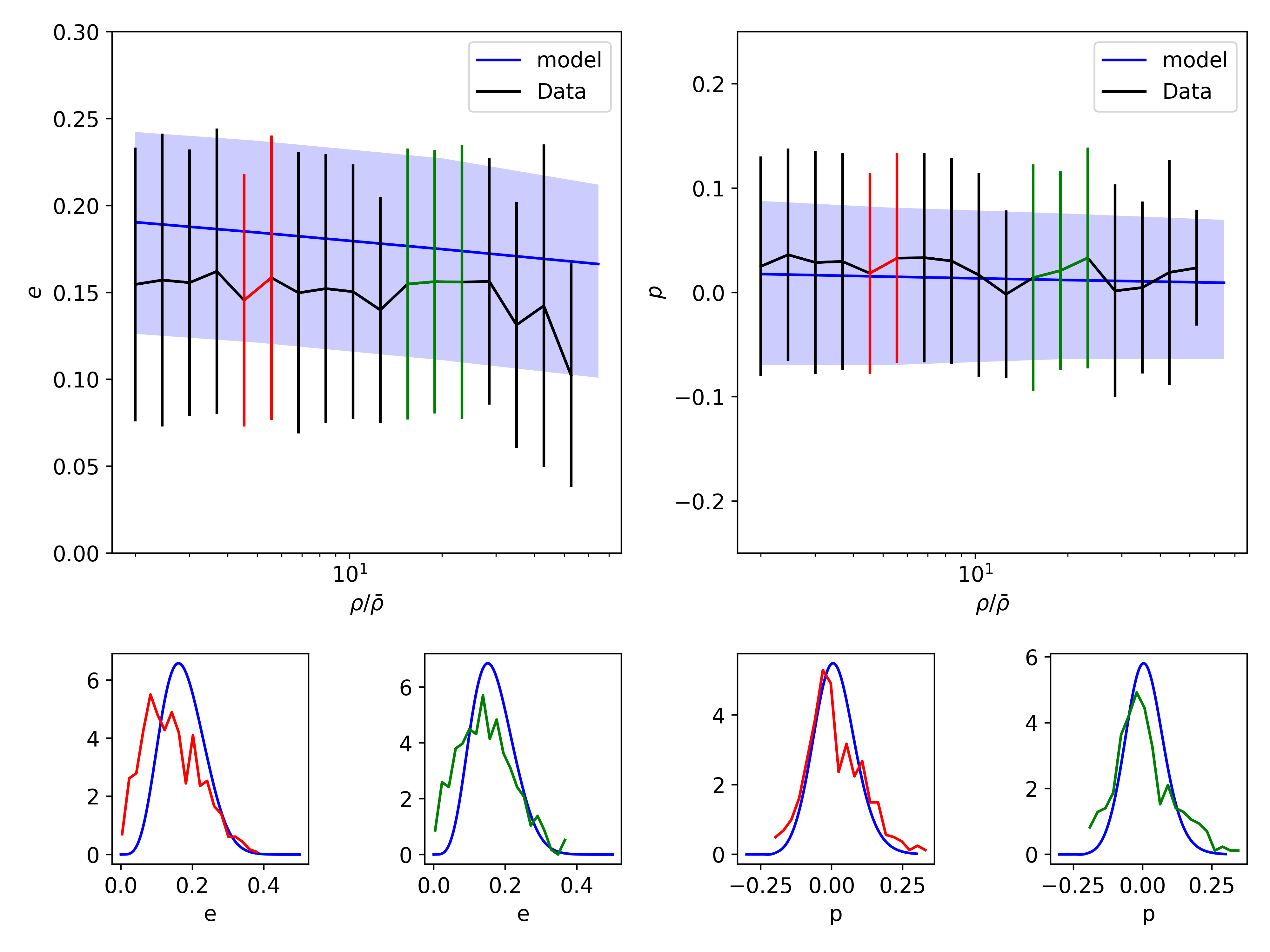}
       \caption{Top: Measured (black) and predicted (blue) distributions of ellipticity (left) and prolateness (right). The structures are extracted with \texttt{Astrodendro} \citep{Robitaille_ExposingPluralNature2019a} in an isothermal turbulent simulation (1024\(^3\)) at Mach 5, with the constraint that the density contour of the perturbations is \(\rho_{\rm max}/\rho_{\rm min}=1.5\) and that each semi-axis is resolved by at least 8 cells. The model is calculated using \(\sigma_s=1.7\) and for structures of size \(R/L_{\rm i}=0.1\), which corresponds to the average size of the structures measured in the simulation. Bottom: Comparison between the predicted distribution (blue) for ellipticity and prolateness for the measured one two different densities (red and green). The density bins that were used to compute this distribution are shown in red and green respectively in the top panel. 
               }
          \label{Fig_numerical_verification}
\end{figure*}

In top of Fig. \ref{Fig_numerical_verification}, we show in black the averaged ellipticity and prolateness. The error bars correspond to the 1\(\sigma\) deviation of the distribution. The model of the evolution of these quantities with the average density of the structure is shown in blue. The shaded area also corresponds to the 1\(\sigma\) deviation of the predicted distribution. We see that our model reproduces well the mean of the distribution and its standard deviation, both for ellipticity and  prolateness. The decrease in deformation predicted by the model as the density of the structure is increased is reasonably well found in the simulation, while slightly overestimated by the model. The standard deviation is slightly underestimated. This is probably a consequence of the Gaussian random field assumption. High order correlations may be responsible for increasing the diversity of the realisations.
Considering that the typical scale \(R\) of the structure is given by the largest axis, the size of the extracted structures varies between \(R/L_{\rm i}=0.05\) and \(R/L_{\rm i}=0.2\). Noting that our model is weakly dependent on the size of the structure, as shown in Fig. \ref{Plot_Shape_distribution}, we compute the model for the averaged extracted size \(R/L_{\rm i}\simeq 0.1\). We have tested that considering a smaller or larger scale by a factor of 2 changes the prediction by at most a few percent. 

In the bottom of Fig. \ref{Fig_numerical_verification}, we directly compare the marginal distribution of \(e\) and \(p\) with the one measured in the simulation for two different densities. We see that they are in reasonable agreement. The peak of the distribution is slightly smaller than that predicted by the model. We note that the agreement between the model prediction and the simulation measurements is still very good, considering that the model has no free parameters.

The success of the numerical verification of the model suggests that the assumption of gaussianity for the logdensity field, while inherently imprecise, is relevant enough to describe the geometry of the density structures that form in the turbulent flow. To test the model over a wider range of densities would require performing the same analysis in a much larger simulation, typically with a resolution close to that of \cite{Federrath_SonicScaleInterstellar2021}, which is beyond the scope of the present study. 

\section{The CMF of an ellipsoidal collapse}
\label{CMF_ellipsoidal_collapse}

To evaluate the effect of the ellipsoidal collapse on the CMF, we proceed as in HC08. We first determine the virial barrier before calculating the CMF.

The statistical description of the ellipticity and prolateness of a density peak applies to any gaussian random field, as pointed out by \cite{Bardeen_StatisticsPeaksGaussian1986}. Before applying this formalism to star formation in the ISM using the HC08 framework, we first justify why the assumption that the logdensity field \(s\) is a gaussian random field is relevant. As will be emphasised below, this assumption is certainly incomplete, but it is a first step in exploring the deformation of the structures formed in the interstellar medium.

\subsection{The gaussian random field assumption}
\label{Sec_gaussian}

\subsubsection{Physical arguments supporting the gaussian assumption}

Strictly speaking, there is no mathematical justification for the Gaussian nature of the logdensity field, even though one can accept the fact that the \(s(\bm{q}_i)\) (logdensity at a point $\bm{q}_i$ of the cloud) are gaussian random variables. This assumption is also at the heart of HC08 and H12\footnote{Note that Hopkins' excursion set approach is based on Brownian random walks, implying (without justification) that the field \(s\) is gaussian.} theories and its validity can only be assessed by comparing the results of these theories with observations or simulations. And indeed, these theories have been shown to be quite successful in explaining the shape and rather universal character of the core mass function (CMF) and of the giant molecular cloud (GMC) mass function over a wide range of cluster conditions in different environments.



Furthermore, observations and simulations of the {\it statistically homogeneous} fluctuating density field exhibit an initial lognormal distribution when smoothed by a window function corresponding to the observational or numerical resolution limit. 
In particular, it is well established that the two-dimensional (2D) column density PDF exhibits LN behaviour, at least in the early stages of molecular cloud evolution when non-self-gravitating supersonic, roughly isothermal turbulence dominates (e.g. \cite{Lombardi_2MASSWideField2010}). 
This log-normality of the 2D PDF translates into the one of the three-dimensional (3D) PDF \citep{Brunt_MethodReconstructingVariance2010}. These results are confirmed by numerous hydrodynamic or magnetohydrodynamic simulations of non-self-gravitating compressible isothermal turbulence (e.g. \cite{Molina_DensityVarianceMachNumber2012}) for the typical Mach numbers characteristic of star-forming molecular clouds. 
Consequently, the PDF of the logarithmic density field $s^{(R)}= \log (\rho_R/{\bar \rho})$ is assumed to be gaussian at all scales. Note that the fact that \(s^{(R)}=\log (\rho \ast W_R/{\bar \rho})\) is gaussian at all scales is not equivalent to saying that \(s_R=s\ast W_R=\log (\rho/{\bar \rho}) \ast W_R\) is gaussian, which would be a strong argument for the gaussian nature of the smoothed logdensity random field. However, given the universality of all observations, we think it is a reasonable assumption. 

From a physical point of view it is worth noting that, as shown below, the typical density increase in the early stages of prestellar clump formation is of the order of \(\sim 10^6/10^2=10^4\), i.e. a factor of a few in logdensity. According to the Rankine-Hugoniot relation, \(\rho_2=\mathcal{M}^2\rho_1\), where \(\rho_1\) and \(\rho_2\) denote the pre- and post-shock densities respectively, this implies about 3-4 successive shocks, for typical Mach numbers in star-forming clouds, \(\mathcal{M}=3-8\).
With such a small number of shocks and small fluctuations in \(s\), the non-linearity of the underlying $s$-field should not be severe. Although this argument is by no means a proof, it provides a plausible explanation for the fact that the gaussian approximation for \(s\) seems reasonable. On the other hand, the fact that the overdensities are generated by multiple shocks suggests that the assumption of {\it local} isotropy for the turbulent flow at the core scale is reasonable. 
Indeed, there is no strong reason why one shock direction should be statistically strongly favoured over the others. If the density fluctuations were the result of only 1 or 2 shocks, the turbulent flow would be anisotropic and the statistical description presented here would be questionable. This is certainly the case for larger scale structures in the ISM, e.g. sheets or filaments, but the present study aims at studying the (de)formation of prestellar {\it cores}, i.e. the statistics of the largest, small scale density fluctuations. 

We therefore admit that {\it at early times} in the star formation process (i.e. before gravity modifies the statistical properties of the turbulence) we are close to gaussian statistics for the logdensity field $s$, allowing us to use the \cite{Bardeen_StatisticsPeaksGaussian1986} approach. 
It may be worth noting in passing that in the Press-Schechter formalism itself, the heuristic derivation of the mass function bypasses all the complications associated with the highly non-linear dynamics of gravitational collapse and clustering. Somehow the same kind of heuristic approach is used in the present calculations.

\subsubsection{Test of the gaussian field assumption: the example of the Polaris and Chameleon clouds}
\label{Sec_test_Polaris}
    
\begin{figure}
    \centering
    \includegraphics[width=\hsize]{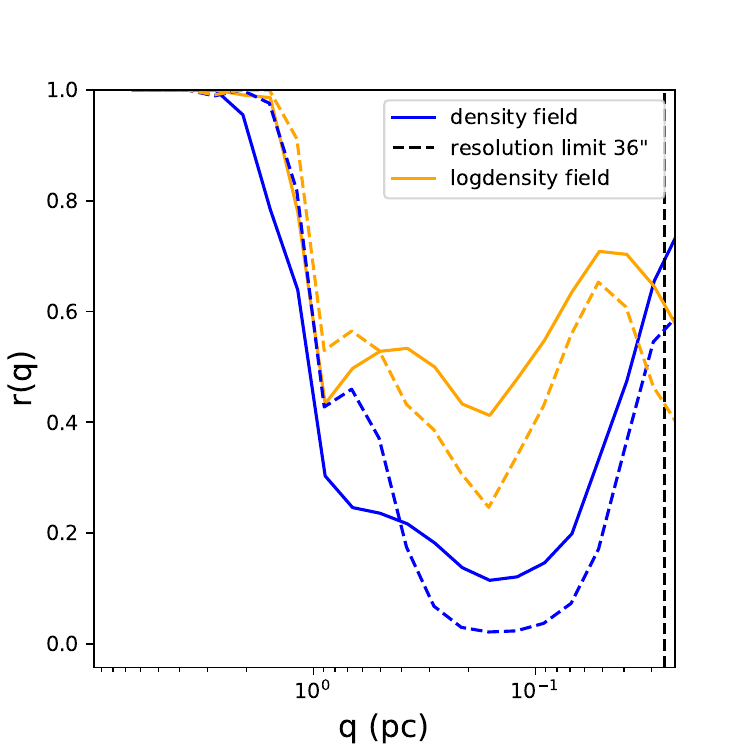}
       \caption{
           Ratio of the gaussian component to total power spectrum (gaussian + coherent) of the density (blue) and logdensity (orange). 
           The wavelet decomposition of the Polaris cloud is shown in solid and the one of the Chameleon III cloud in dash.
               }
          \label{Fig_wavelet}
\end{figure}

As mentioned above, the assumption that the logdensity field is gaussian is equivalent to the assumption that all statistical information is contained in the power spectrum. One way to quantify the relevance of this assumption is to perform a wavelet decomposition of the logdensity field to extract the gaussian and non-gaussian features separately and compute their respective power. This has been done in observations and simulations by \cite{Robitaille_ExposingPluralNature2019} and \cite{Colman_SignatureLargeScale2022}, respectively. In Fig. \ref{Fig_wavelet}, we plot the ratio of the power spectrum of the gaussian component to the total power spectrum (gaussian + coherent) for the {\it logarithmic density} of the Polaris (solid line) and Chameleon III (dashed line) clouds \citep{Andre_FilamentaryCloudsPrestellar2010}. These two clouds were chosen because they are  representative of the early stages of star formation, i.e. before gravity dominates the dynamics of density perturbations formation, and their density PDF is still close to lognormal.
We also compare the same ratio for the {\it density}, as already done by \cite{Robitaille_ExposingPluralNature2019} (their fig. 14). 

While the gaussian assumption is incorrect for the density field, a well-known result, the gaussian contribution to the logdensity field is dominant in the ranges $\sim 1$-10 pc and $\lesssim$0.2 pc. In the HC08 framework, the latter scale corresponds to the formation of a prestellar core with a mass of the order of 15 M\(_\odot\). This analysis of the observational data shows that, at first order, our assumption of gaussian statistics for the logdensity field of the initial {\it prestellar core} density fluctuations is a fairly reasonable approach to an analytical treatment of the problem of the shape of density perturbations. In the intermediate scale region, the non-gaussian contribution is no longer negligible and may be comparable to, or even dominant over, the gaussian contribution. The fact that we may potentially underestimate the deformation of some density perturbations due to the non-negligible role of non-Gaussianities is discussed in more detail in the section \ref{Sec_core_mass_function}.

We see that the non-gaussian contribution is strongest for scales around 0.2 pc, close to the typical width of star-forming filaments \citep{Federrath_UniversalityInterstellarFilaments2016, Arzoumanian_CharacterizingPropertiesNearby2019}, for both the density and the logdensity field. This means that while non-gaussian statistical processes for turbulent induced density fluctuations are not negligible for filament formation, the gaussian processes become dominant at smaller scales, i.e. at the core scales, the domain of interest in the present study.

\subsection{The virial barrier}

At scale $R$, the critical overdensity above which the ellipsoidal perturbation will collapse is given by the virial condition:
\begin{equation}
    \langle V_{\rm RMS}(R)^2\rangle+3C_{\rm s}^2<-\frac{E_{\rm p}(R)}{M},
     \label{virial}
\end{equation}
where \(C_{\rm s}=0.2\,\text{km/s}\) is the sound speed for a medium at \(T=10\text{K}\), \(E_{\rm p}\) the gravitational energy of an ellipsoid of mass \(M\) and $V_{\rm RMS}$ the turbulent velocity dispersion, which follows the Larson relation \citep{Larson_TurbulenceStarFormation1981}:
\begin{equation}
    V_{\rm RMS}^2(R)=V_0^2\left(\frac{R}{1\text{pc}}\right)^{2\eta},
\end{equation} 
with \(\eta\simeq 0.5\) \Citep{Falgarone_StructureMolecularClouds2004}, \(V_0\simeq0.8 \text{km/s}\).
As we consider only the central part of the perturbation close to the maximum density $ \rho_0$, we assume that the ellipsoid is homogeneous. The mass of such an ellipsoid is \(M=4\pi/3\bar{\rho}\e^{s}a_1 a_2 a_3\), where the \(a_i\)'s are the semi-axis of the ellipsoid given by eq. (\ref{semi_axis}), with \(a_3>a_2>a_1\). Its gravitational energy is given by (\cite{Chandrasekhar_EllipsoidalFiguresEquilibrium1969}):

\begin{equation}
    E_{\rm p} = -\frac{2}{5}\pi GM\rho a_2^2 I,
\end{equation}
where 
\begin{equation}
    I(a_1, a_2, a_3)=\int_0^\infty\frac{du}{\sqrt{(1+\left(\frac{a_2}{a_1}\right)^2u)(1+\left(\frac{a_2}{a_3}\right)^2u)(1+u)}}.
\end{equation} 
In the case of a sphere of radius $R$, all the eigenvalues of the deformation tensor are equal, which yields the well known result $E_{\rm p}^{\rm sph}=-(2/5)\pi GM\rho I=-(3/5) GM^2/R$, with $I=2R^2$.

\noindent At fixed mass and density, the most unstable structure is the spherical one. Indeed, \(\forall (a_1, a_2), I(a_1, a_2, 1/a_1)<2\) and reaches the equality for \(a_1=1\) which corresponds to a sphere, meaning that \(E_{\rm p}\) is increased when the structure is deformed. The choice \(a_3=1/a_1\) ensures that the mass is constant for a given density and \(a_2\). Therefore, the more deformed a structure, the higher its gravitational energy. 

\noindent As mentioned in \S\ref{Sec_gaussian}, we proceed exactly as in HC08. For an ellipsoid  formed at scale $R$ characterized by eigenvalues $\lambda_1, \lambda_2, \lambda_3$, the  threshold density  for collapse can be expressed as:
\begin{eqnarray}
    \label{Virial_barrier}
    \frac{\rho_R^{\rm c}}{\bar{\rho}}=\frac{2}{I}\frac{1+\mathcal{M}_\star^2(2\tilde{a}_3)^{2\eta}}{\tilde{a}_2^2},
  \end{eqnarray}
  to be compared with equation (29) of HC08 for the spherical collapse condition, where $\mathcal{M}_\star$ defines the Mach number at the Jeans scale \(\lambda_{\rm J}=C_{\rm s}/\sqrt{G\bar{\rho}}\) (see HC08):
\begin{equation}
    \mathcal{M}_\star^2=\frac{1}{3}\left(\frac{V_0}{C_{\rm s}}\right)^2\left(\frac{\lambda_{\rm J}}{1\text{pc}}\right)^{2\eta}.
\end{equation}
  Taking the scale $R$ as the smallest semi-major axis of the ellipsoid, the associated critical mass is:
  \begin{eqnarray}
    M_R^{\rm c}&=&  \frac{4\pi}{3}{\bar{\rho}}{a_3^3}\sqrt {\alpha \beta} \,e^{s_R^c},
  \label{mcrit}
\end{eqnarray}
where $\alpha=\lambda_1/\lambda_3$, $\beta=\lambda_2/\lambda_3$ and the axis are normalized to the Jeans length, $\tilde{a}_i={a}_i/\lambda_{\rm J}$. With our ordering convention of the eigenvalues, we have \(\alpha>\beta>1\).
In Appendix \ref{Rotation}, we show than the contribution of the rotational energy of the prestellar clumps is negligible compared with the one from turbulence and is almost scale-independent. It will thus not impact the CMF.

As seen in eq.~(\ref{Virial_barrier}), the ellipsoid collapse condition involves the integral $I$, so the collapse barrier now depends on the shape of the ellipsoid, through its eigenvalues. 
The non-sphericity of the perturbation {\it increases} the collapse threshold density compared with the spherical case. This is similar to the collapse of dark matter halos in cosmology \citep{Sheth_EllipsoidalCollapseImproved2001}. We will come back to this point later (see Section \ref{cosmo}).
As mentioned above and seen in Fig. \ref{Plot_Shape_distribution}, it is interesting to note that these terms depend very weakly on the scale. 
The virial barrier will still be slightly more increased for the low density structures (formed at large scales) than for the densest ones.
We also note from eq.~(\ref{mcrit}) that, at a given scale, the bound structure contains more mass than in the spherical case.

\subsection{The Core Mass Function}
\label{Sec_core_mass_function}

An exact calculation of the CMF would require to consider the barrier for each possible ellipsoid, a numerically quite heavy task. To keep the calculations reasonable, we consider only the {\it average barrier} corresponding to the average ellipsoid formed at scale \(R\) with a logdensity contrast \(s_R\) defined as:
\begin{equation}
    \langle s_R^{\rm c}\rangle(s)=\int_{0}^{\infty}\mathcal{P}_R(\bm{\lambda}|s)s_R^{\rm c}(\bm{\lambda}){\rm d}^3\bm{\lambda},
\end{equation}
where \(s_R^{\rm c}\) is defined from eq. \ref{Virial_barrier} as \(s_R^{\rm c}=\ln(\rho_R^{\rm c}/\bar{\rho})\). \(\mathcal{P}_R(\bm{\lambda}|s)\) is the PDF of the random vector \(\bm{\lambda}=(\lambda_1, \lambda_2, \lambda_3)\) knowing the logdensity \(s\) of the ellipsoid.

Strictly speaking, the density of a structure formed at the scale \(R\) is not \(\bar{\rho}\e^{s_R}=\bar{\rho}\exp(s\ast W_R)\) but $\rho_R=\bar{\rho}e^{s^{(R)}}=\rho\ast W_R$, as \(s_R\ne s^{(R)}\). However, the exponential is a strictly convex function, and from the convexity inequality, if \(s_R=\ln(\rho_R^{\rm c}/\bar{\rho})\), then \(\rho_R>\rho_R^{\rm c}\). Thus, the necessary and sufficient collapse condition on \(\rho_R\) (eq. \ref{Virial_barrier}) is only a sufficient collapse condition on \(s_R\). As a consequence, this condition underestimates the number of unstable regions, since some of them could satisfy the condition \(\rho_R\ge\rho_R^{\rm c}\), but with \(s_R< \ln(\rho_R^{\rm c}/\bar{\rho})\). However, from the point of view of a {\it statistical} collapse criterion based on a very large number of density fluctuations, it seems reasonable to consider \(s_R^{\rm c}\) as the collapse criterion.

\begin{figure*}
    \centering
    \includegraphics[width=\hsize]{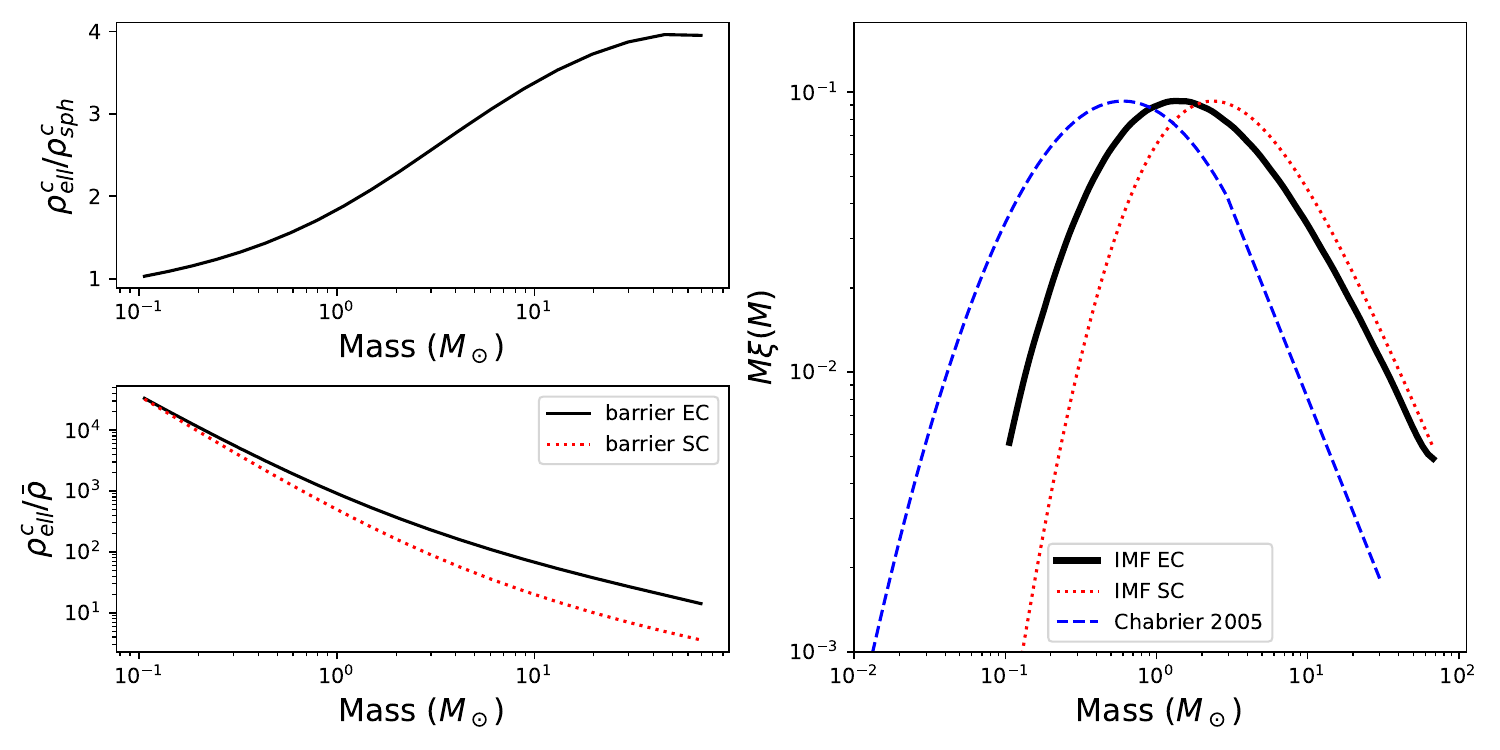}
       \caption{Left: virial barrier in the case of ellipsoidal (EC) and spherical (SC) collapse (bottom) and ratio of these two barriers (top). Right: CMF in the case of ellipsoidal and spherical collapse. The dashed line corresponds to the Chabrier (2005)  IMF shifted to higher masses by a star-to-core uniform factor $\sim$3 for the CMF (see HC08 sec. 7.1.2). The injection scale for turbulence is chosen to be \(L_{\rm i}=10\text{pc}\). }
        \label{fig_CMF}
\end{figure*}

Thus, the ellipsoids that are virialised on average are those that satisfy the conditions \(s=\langle s_R^{\rm c}\rangle(s)\). By solving this equation at each scale, we can derive the average virial barrier of the system. As the shape of the ellipsoid depends only weakly on the density of the perturbation, as mentioned above, the barrier is found to be qualitatively very similar to that obtained in HC08 for spherical collapse. It is shifted to higher densities by the geometric factor \(2/I\) and is slightly higher for the high masses than for the low masses, since the former ones are the most susceptible to deformation. This is illustrated in the left panel of Fig. \ref{fig_CMF}, where we see that the ratio of the ellipsoidal to the spherical barrier increases by a factor of $\sim$3 when the mass is increased by two orders of magnitude.
Consequently, we expect the CMF of ellipsoid bound cores to differ from that obtained for spherical collapse by a larger relative number of high mass objects. This is explored below.

To compute the prestellar core CMF, \(\xi(M)=dN/dM\), we proceed as in HC08. Within this formalism, the mass enclosed in bound cores is obtained from 2 conditions, which insures that the cloud-in-cloud problem is resolved.
The first condition reads: 
\begin{equation}
    \label{first_condition}
    M_{\rm tot}(R)=\int_{s_{R}^c}^{\infty} L_{\rm i}^3\mathcal{P}_R(s) \int_{\mathbb{R}_+^3} \mathcal{P}_R(\bm{\lambda} | s)M(s, \bm{\lambda}){\rm d}\bm{\lambda}{\rm d}s,
\end{equation}
where the second integral is the average mass contained in a clump formed at scale \(R\) with a logdensity contrast \(s\) and \(\mathcal{P}_R(s_R^{\rm c})\) is the marginal distribution of the logdensity peaks calculated from equation \ref{Joint_g_x_e_p} and can be found in \cite[Appendix A]{Bardeen_StatisticsPeaksGaussian1986}. Multiplying by the  number of clumps formed under these conditions (first integral), we obtain the total mass gathered in clumps formed at scale \(R\) whose
density exceeds the  threshold density for collapse. The resulting cores will collapse, and we consider that these cores will not exceed the density threshold at  scale \(R\) because denser structures will fragment. On the other hand, this total mass will be distributed in cores with a mass smaller than the average mass threshold \(M_R^{\rm c}\):
\begin{equation}
     M_{\rm tot}(R)=\int_0^{M_R^{\rm c}}L_{\rm i}^3 M'\xi(M'){\rm d}M'.
\end{equation}
By equating the two expressions and neglecting the second integral in the derivation of these quantities, which is a valid assumption for scales \(R\ll L_{\rm i}\) (see Appendix B of HC08), we get
\begin{equation}
    \label{eq_CMF}
    M_R^{\rm c} \xi(M_R^{\rm c})=-\frac{dR}{dM_R^{\rm c}}\frac{ds_R^{\rm c}}{dR}\mathcal{P}_R(s_R^{\rm c})M_R^{\rm c},
\end{equation}
where the average critical mass is given by
\begin{equation}
    M_R^{\rm c}=\int_{\mathbb{R}_+^3} \mathcal{P}_R(\bm{\lambda} | s_R^{\rm c})M(s_R^{\rm c}, \bm{\lambda}) {\rm d}\bm{\lambda}.
\end{equation}

The resulting CMF is plotted in the right panel of Fig. \ref{fig_CMF}. The turbulence is injected at \(L_{\rm i}=10\,\text{pc}\) and is dissipated at scales smaller than 0.01pc, in agreement with observations of the density power spectrum \citep{Miville-Deschenes_ProbingInterstellarTurbulence2016,Pineda_ProbingPhysicsStarFormation2024}. This value is indeed negligible relative to the injection scale as assumed in Appendix \ref{App_Correlation_function}, implying that the derived CMF will not depend on its true value.
The plotted CMF in Fig. \ref{fig_CMF} corresponds to structures formed in a cloud whose background density is given by the first Larson condition \(\bar{n}=3500(L_{\rm i}/1\text{pc})^{-0.7}\text{cm}^{-3}\). These conditions are just illustrative. We present an example of a CMF obtained with a larger injection scale in appendix \ref{Other_injection_lengths}.
As mentioned above, the present calculations aim at characterizing the impact of the departure from sphericity for collapsing prestellar cores. Thermodynamic effects as explored in \cite{Hennebelle_ANALYTICALTHEORYINITIAL2009}, which increase the relative number of small-scale structures, are not considered presently.

As can be seen in Fig. \ref{fig_CMF}, and as expected from the above study of the distribution of ellipsoid shapes and collapse state, the CMF for ellipsoidal collapse is similar to that obtained for spherical collapse, with a slightly flatter slope for the large mass range, even more evident for larger injection scales (see Fig. \ref{fig_CMF_20}). This again contrasts with the cosmological case, a result we discuss below. Since the ratio \(\rho^{\rm c}_{\rm ell}/\rho^{\rm c}_{\rm sph}\) is essentially scale independent for low-mass cores, the small-mass range of the CMF is less affected by ellipsoidal collapse than the high-mass range. This results in a small shift of the peak towards small masses compared to spherical collapse, as seen in the figure.

It should be noted that the present study has two limitations. First, we consider the density and velocity fields to be decorrelated and thus described by independent statistics (as in HC08 and H12); second, we ignore the effect of the magnetic field. We plan to explore these effects in more detail in future studies. However, these approximations seem reasonable. First, while for transonic turbulence density and velocity are significantly correlated, in the supersonic regime the correlations tend to weaken with increasing Mach number \citep{Rabatin_DensityVelocityCorrelations2023}. Indeed, simulations show that in isothermal supersonic turbulence $v$ is weakly correlated with $\rho$ (see e.g. Fig. 7 of \citealt{Federrath_DensityStructureStar2015}).
As for the magnetic field, it affects the structure of the turbulent flow, which can become highly anisotropic for strong magnetic fields (Alfvénic Mach number ${\cal M}_{\rm A}=V_{\rm RMS}/v_{\rm A}\ll 1$, where $v_{\rm A}$ is the Alfvén mean velocity). However, for star-forming clouds ${\cal M}_{\rm A}\sim 1$ and the effect of the magnetic field on the PDF remains modest (e.g., \citealt{Molina_DensityVarianceMachNumber2012, Beattie_MultishockModelDensity2021}), slightly decreasing the variance. The anisotropy of the turbulent flow could induce a larger deformation of the structures. This will be investigated in a future study.

So far we have not considered the dynamics of the deformation. However, we do not expect the results to change significantly qualitatively in this case. Indeed, the structures most affected by the dynamics, i.e. those with the shortest collapse time, are the densest, smallest mass ones \citep{Hennebelle_ANALYTICALSTARFORMATION2011,Hennebelle_ANALYTICALTHEORYINITIAL2013}, which are the less prone to deformation. 
We examine the effect of time dependence on the CMF for ellipsoidal collapse in section \ref{Time_dependence}.

In section \ref{Sec_test_Polaris}, we showed that while non-Gaussianities are not dominant at any scale, they play a non-negligible role in a certain range ($\sim 0.1$-1 pc). This could lead to an underestimation of the deformation of the density perturbations at these scales. In particular, while we predicted that the maximum aspect ratio of the structures is of the order of 2, which is probably true at small (core) scales, small-scale filaments, i.e. fibres, exhibit aspect ratios of up to 10 (see e.g. \cite{Hacar_FibersNGC13332017a, Hacar_InitialConditionsStar2022}). To test this effect, we artificially increased the predicted deformation by a factor of 4. This does not change the qualitative result of our study, i.e. that only the high mass range is affected, not the low mass range. Quantitatively, it makes the high-mass slope slightly flatter, with a slope index increased by about 15\%, still consistent with observations. 

\subsection{Time dependence}
\label{Time_dependence}

\begin{figure}
	\centering
	\includegraphics[width=\hsize]{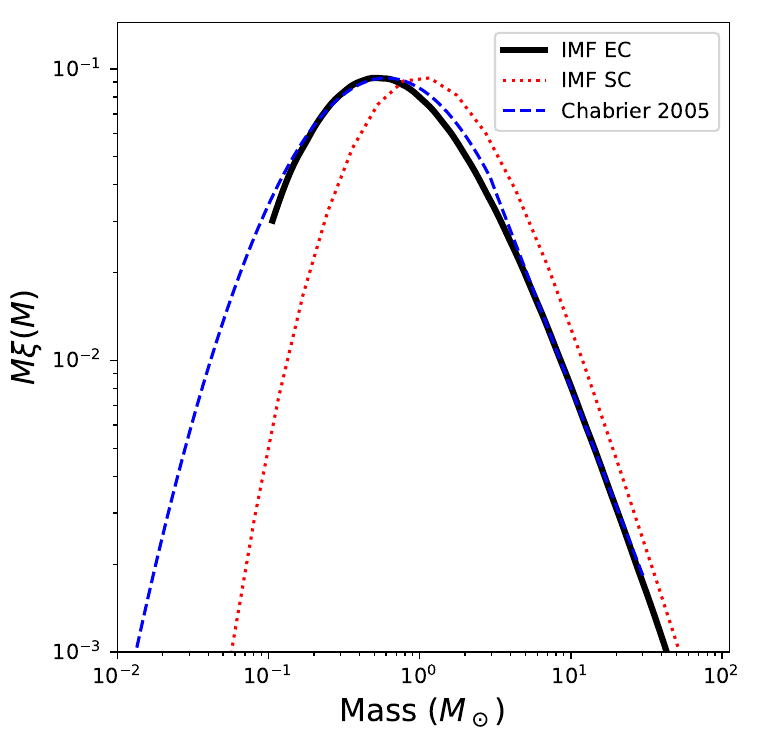}
	   \caption{Similar as the right panel of Fig. \ref{fig_CMF} taking into account the time dependence for the spherical and ellipsoidal model and for a turbulent injection scale \(L_{\rm i}=10\)pc.}
		\label{CMF_time_dependence}
\end{figure}

Here, we study the impact of the time dependence of the formation of turbulent perturbations as in \cite{Hennebelle_ANALYTICALSTARFORMATION2011,Hennebelle_ANALYTICALTHEORYINITIAL2013}. At each scale, perturbations are formed at each new flow of turbulence. The rate of new turbulent large scale flows can be estimated as \(\tau_{\rm ff}^0/\tau_R\propto \sqrt{\rho/\rbar}\), where \(\tau_{\rm ff}^0\) is the free-fall time of the parent molecular cloud, of the order of its lifetime, and \(\tau_R = R/V_{\rm RMS}(R)\) is the crossing time of turbulence. Thus, the mass enclosed in bound cores (eq.~\ref{first_condition}) becomes:
\begin{equation}
	M_{\rm tot}(R)\propto\int_{s_{R}^c}^{\infty} L_{\rm i}^3\mathcal{P}(s) \e^{s/2}\int_{\mathbb{R}_+^3}  \mathcal{P}_R(\bm{\lambda} | s)M(s, \bm{\lambda}){\rm d} \bm{\lambda}{\rm d}s.
\end{equation}
Then, the expression of the CMF taking into account the number of turbulence injection flows is:
\begin{equation}
    \label{eq_CMF_time}
	M_R^{\rm c} \xi(M_R^{\rm c})\propto-\frac{{\rm d}R}{{\rm d}M_R^{\rm c}}\frac{{\rm d}s_R^{\rm c}}{{\rm d}R}\mathcal{P}(s_R^{\rm c})\e^{s_R^{\rm c}/2}M_R^{\rm c}.
\end{equation}
It is plotted in Fig. \ref{CMF_time_dependence}. As found in  \cite{Hennebelle_ANALYTICALSTARFORMATION2011,Hennebelle_ANALYTICALTHEORYINITIAL2013}, we recover a slope for the high mass tail very close to the Salpeter one and the peak of the predicted CMF is consistent with the observed one. Our prediction matches even better the observed CMF.

We can estimate the modification of the slope of the CMF for the high mass range. We should first emphasize that the PDFs used in the spherical collapse model and here are not the same. The PDF used in HC08 is purely lognormal whereas the one in ellipsoidal collapse model takes into account the correlation between the density and the second order moments. Moreover, the PDF used here is the number of density maxima per unit volume which modifies the expression of the CMF (eq. \ref{eq_CMF_time}) with respect to the one given in HC08 (their equation (33) without the second integral). To explicit each term involved in the expression in the CMF and important for the Salpeter slope, we can write eq. \ref{eq_CMF_time} more explicitly as:  
\begin{equation}
	M \xi(M)\propto-\frac{{\rm d}s_R^{\rm c}}{{\rm d}M}\e^{-(s_R^{\rm c}+\sigma_s^2/2)^2/2\sigma_s^2}\e^{s_R^{\rm c}/2}M G_R(s_R^{\rm c})\left(\frac{\sigma_\eta(R)}{\sigma_\mathbb{I}(R)}\right)^3,
\end{equation}
where \(G(s_R^{\rm c})\) is a function expressing the correlation between the density and the second order moments.
The three last terms scale as a density, and we can see numerically that \(M G_R(s_R^{\rm c})\left(\frac{\sigma_\eta(R)}{\sigma_\mathbb{I}(R)}\right)^3\propto \bar{\rho}\e^{\kappa s_R^{\rm c}}\) with \(\kappa\) close to 1 (the found value is \(\kappa \simeq 1.15\)). Thus, the present expression of the CMF behaves similarly as the one given in HC08. The observed differences in the shape of the CMF is thus mainly due to the modification of the barrier.

The dependence of the barrier on the mass changes from \(\rho_R^{\rm c}/\bar{\rho} = \beta \tilde{M}^{\alpha}\) to \(\beta' \tilde{M}^{\alpha+\gamma}\) where \(\alpha<0\), \(\gamma>0\), \(\beta'>\beta\) and \(\tilde{M}\) is the mass normalized to the Jeans mass \(M_{\rm J}=4\pi/3\lambda_{\rm J}^3\bar{\rho}\). We can estimate the slope \(y\) of the CMF in the time dependent case:
\begin{equation}
	y=\alpha+\gamma-1-\frac{\alpha+\gamma}{2\sigma_s^2}\ln(\beta')-\frac{(\alpha+\gamma)^2}{2\sigma_s^2}\ln(\tilde{M}).
\end{equation}
We can neglect the last term which does not contribute much in the mass range as we are close to the Jeans mass. With \(\beta'=40, \beta'/\beta=4, \gamma=0.1, \sigma_s^2=3\) and \(\alpha=-1\), as found numerically, we can thus estimate a difference of 0.3 between the two slopes, which is of the order of what is found in fig. \ref{CMF_time_dependence}. Thanks to this estimation, we can see that the correction on the high mass tail is more due to the enhancement of the barrier in the ellipsoidal case than to the modification of its slope.

\section{Spherical collapse: differences between the cosmological and  stellar cases}
\label{cosmo}

As mentioned briefly in the above sections, there are several noticeable differences between the cosmological and the star formation ellipsoidal collapse. We examine below these differences.

A first difference is that in the cosmological case, the perturbations are induced by the initial {\it shear (tidal) field}  (e.g. \citealt{Bond_PeakPatchPictureCosmic1996,Sheth_EllipsoidalCollapseImproved2001,Ludlow_FormationCDMHaloes2014}). As shown by \cite{Ludlow_FormationCDMHaloes2014}, the three eigenvalues in that case are not related to the three axes of the ellipsoid but to the amplitude of the deformation due to the tidal field. The initial conditions are chosen to recover the Zeldovich approximation in the linear regime and virilization occurs when the {\it three} axes of the spheroid collapse, while  the density is given by the Poisson equation. 
Therefore, the collapse barrier in cosmological fluctuations is determined by considering the dynamic of each axis and the time needed for them to collapse. At a given scale, the more deformed the axes of the density fluctuation, the longer the collapsing time (e.g., \citealt{White_GrowthAsphericalStructure1979}). Then, for the same amount of mass in the collapsing structures at a given redshift, distorted structures must have higher density contrasts than spherical ones to fulfil the collapse condition, which translates into a higher barrier \citep{Sheth_EllipsoidalCollapseImproved2001}. The cosmological approach does not concern the formation of an ellipsoid per se but rather its deformation by a tidal field.

In the stellar case, such a simple approach is not possible because the sum of the eigenvalues of the deformation tensor \(\mathbb{I}_{ij}\) cannot be linked to the density contrast of the perturbation through the Poisson equation. This is due to the fact that the relevant field to consider in the stellar case is the logdensity field (which can be assumed to be a gaussian random field) and not the density field. We would need to consider the effect of an external massive object added by hand to the perturbation, which would impose a non-uniform background. 
Furthermore, considering tidal deformation would be inconsistent with a lognormal density field of fluctuations. In that case, the PDF must include the gravity induced power law tails \citep{Federrath_STARFORMATIONEFFICIENCY2013, Burkhart_StarFormationRate2018,Jaupart_EvolutionDensityPDF2020,Khullar_DensityStructureSupersonic2021}, and must take into account the correlation between the velocity and the density fields. This will be examined in a forthcoming study (Dumond \& Chabrier in prep.).
The deformation of a density fluctuation in the present stellar case is thus only of geometrical nature and corresponds to the deformations induced by isotropic turbulence in the initial density field, not by gravity, as for tidal deformations; it  does not involve the presence of an external body. In that case collapse occurs under the action of gravity alone, when the {\it global} virial condition given by eq.(\ref{virial}) is fulfilled. Since the most distorted structures are the most stable, this yields also a higher barrier in the ellipsoidal collapse for prestellar cores.
Therefore, although both in the stellar and cosmological cases departure from sphericity has the same consequence, namely stabilising the structures and thus increasing the collapse barrier compared with the spherical case, the reasons are different.

Finally,  another major difference is that in the cosmological case, the critical density required for a perturbation to collapse at a given redshift $z$ is approximately independent of the mass \citep{Press_FormationGalaxiesClusters1974}. As a consequence, the objects the most prone to deformation are the smallest ones \citep{Sheth_EllipsoidalCollapseImproved2001} because the ratio of the critical density to the density variance $\rho_R^{\rm c}/\sigma_\rho(R)$ is the smallest at small scale (see Section \ref{Shape_distribution}).
At the opposite, in the stellar case, the ratio $s_R^{\rm c}/\sigma_s(R)$ is the highest for the smallest masses and the lowest for the largest ones. Then, the large scale (mass) fluctuations will be the ones the most affected by non-spherical deformations, as seen in the left panel of Fig. \ref{fig_CMF}. This can be intuitively understood since in both cases, departure from sphericity affects more the low-density than the high-density fluctuations: it is easier to deform less dense objects than very dense ones. However, while in the cosmological case the former ones correspond to small-scale (low-mass) halos, it is the opposite in the stellar case: high mass prestellar density fluctuations are easier to deform. This is what we observed in the probability distributions.

\section{Discussion and Conclusion}
\label{Discussion}

In this paper, we have examined the geometry of the initial density fluctuations that will yield eventually to the formation of prestellar cores  by taking into account the full triaxial nature of turbulence induced perturbations in star-forming clumps. 
We found that the fluctuations have indeed an ellipsoidal shape rather than a spherical one. Statistically speaking, however, the spheroids are found to have a negligible prolateness and an ellipticity varying from about 0.05 to 0.15 at the 95\% confidence level for the most distorted structures. We found that the deformation depends weakly on the scale and the density of the fluctuation, but the less dense structures are the most distorted. These theoretical results are confirmed by comparing to the distribution of the shapes of the structures extracted in a numerical simulation of a turbulent box. As a consequence, this analysis shows that the spherical approximation used in the current theories of the CMF/IMF is reasonable. 
Interestingly, as shown in Sect. \ref{Shape_distribution}, the range of shapes of the perturbations in a turbulent medium is much smaller than in the cosmological case, a consequence of the lognormality of the density field.

We have then examined the impact of the departure from sphericity on the collapse condition for the formation of bound structures and the resulting CMF. We found that the barrier is increased compared to the one obtained for the spherical collapse condition: the stronger the deformation the more stable the perturbation, a direct consequence of the stabilization of the ellipsoid induced by its elongation.  As a consequence of the aforementioned weak scale dependence of the deformation of the density perturbations in the stellar case, the departure from sphericity of the overdense structures in that case affects nearly uniformly the CMF, although it yields a slightly shallower high-mass tail.
This is in contrast with the cosmological case where ellipticity of the primordial density fluctuations is essential to predict the observed halo mass function by suppressing the excess of low mass structures compared with the spherical collapse. This arises essentially from the very different probability laws and power spectrum of fluctuations induced by turbulence compared with the ones triggered by gravitational tidal fields. Interestingly enough, we note that the aforementioned modification of the high-mass tail of the CMF with the ellipsoidal collapse yields a slight shift of the peak toward smaller masses. This improves the agreement with the \cite{Chabrier_InitialMassFunction2005} IMF (see Fig. \ref{fig_CMF_20} and \ref{CMF_time_dependence}) compared to the prediction of the HC theory taking into account the time dependence \citep{Hennebelle_ANALYTICALTHEORYINITIAL2013}.

Another striking difference between the cosmological and stellar cases is that in the former, the ratio \(\rho_R^{\rm c}/\sigma_\rho(R)^2\) decreases with mass, so that small masses are more prone to distortion. In contrast, in the stellar case, large masses are more prone to distortion because \(s_R^{\rm c}/\sigma_s(R)^2\) decreases with {\it increasing} mass. As a consequence, the ellipsoidal barrier will increase more for large scales (masses) than for small ones, compared to the spherical one. The predicted slope of the CMF is therefore slightly flatter in this large mass range. This deviation from the Salpeter slope is corrected by taking into account the time dependence of the formation of fluctuations as in \cite{Hennebelle_ANALYTICALSTARFORMATION2011,Hennebelle_ANALYTICALTHEORYINITIAL2013} (see section \ref{Time_dependence}). The CMF predicted with the dependent ellipsoidal model is in even better agreement with the observations than the spherical model.

We emphasise that the present study represents a first step towards the triaxial collapse of prestellar cores, since we have only considered structures formed by isotropic turbulence, where the density field is modelled by a lognormal random field. We have not considered the effect of gravitational perturbations leading to power-law tails in the PDF. However, as this regime occurs when gravity dominates the dynamics of the system, i.e. in the densest regions of the PDF \citep{Jaupart_EvolutionDensityPDF2020, Jaupart_StatisticalPropertiesCorrelation2022}, we can expect the structures to be less deformed than in the turbulent case. We therefore expect the most gravitationally unstable structures in the gravity dominated regime to be spheroidal.

\section*{Acknowledgements}
We thank the anonymous referee for constructive feedback that helped to improve the quality of this work. We are grateful  to Jeremy Fensch and Guillaume Laibe for their careful reading of the manuscript and helpful comments. We also thank Thomas Gillet for helpful discussions.

\section*{DATA AVAILABILITY}
The data underlying this article can be shared for selected scientific purposes after request to the corresponding author.



\bibliographystyle{mnras}
\bibliography{Bib} 

\begin{thebibliography}{}
\makeatletter
\relax
\def\mn@urlcharsother{\let\do\@makeother \do\$\do\&\do\#\do\^\do\_\do\%\do\~}
\def\mn@doi{\begingroup\mn@urlcharsother \@ifnextchar [ {\mn@doi@}
  {\mn@doi@[]}}
\def\mn@doi@[#1]#2{\def\@tempa{#1}\ifx\@tempa\@empty \href
  {http://dx.doi.org/#2} {doi:#2}\else \href {http://dx.doi.org/#2} {#1}\fi
  \endgroup}
\def\mn@eprint#1#2{\mn@eprint@#1:#2::\@nil}
\def\mn@eprint@arXiv#1{\href {http://arxiv.org/abs/#1} {{\tt arXiv:#1}}}
\def\mn@eprint@dblp#1{\href {http://dblp.uni-trier.de/rec/bibtex/#1.xml}
  {dblp:#1}}
\def\mn@eprint@#1:#2:#3:#4\@nil{\def\@tempa {#1}\def\@tempb {#2}\def\@tempc
  {#3}\ifx \@tempc \@empty \let \@tempc \@tempb \let \@tempb \@tempa \fi \ifx
  \@tempb \@empty \def\@tempb {arXiv}\fi \@ifundefined
  {mn@eprint@\@tempb}{\@tempb:\@tempc}{\expandafter \expandafter \csname
  mn@eprint@\@tempb\endcsname \expandafter{\@tempc}}}

\bibitem[\protect\citeauthoryear{Adler}{Adler}{1981}]{Adler_GeometryRandomFields1981}
Adler R.~J.,  1981, The {Geometry} of {Random} {Fields}.
\url {https://ui.adsabs.harvard.edu/abs/1981grf..book.....A}

\bibitem[\protect\citeauthoryear{Andre, Di~Francesco, Ward-Thompson, Inutsuka,
  Pudritz  \& Pineda}{Andre et~al.}{2014}]{Andre_FilamentaryNetworksDense2014}
Andre P.,  Di~Francesco J.,  Ward-Thompson D.,  Inutsuka S.-i.,  Pudritz R.~E.,
    Pineda J.,  2014, From {Filamentary} {Networks} to {Dense} {Cores} in
  {Molecular} {Clouds}: {Toward} a {New} {Paradigm} for {Star} {Formation},
  \mn@doi{10.2458/azu_uapress_9780816531240-ch002}, \url
  {http://arxiv.org/abs/1312.6232}

\bibitem[\protect\citeauthoryear{André et~al.,}{André
  et~al.}{2010}]{Andre_FilamentaryCloudsPrestellar2010}
André P.,  et~al., 2010, \mn@doi [Astronomy and Astrophysics]
  {10.1051/0004-6361/201014666}, 518, L102

\bibitem[\protect\citeauthoryear{Arzoumanian et~al.,}{Arzoumanian
  et~al.}{2019}]{Arzoumanian_CharacterizingPropertiesNearby2019}
Arzoumanian D.,  et~al., 2019, \mn@doi [Astronomy \& Astrophysics]
  {10.1051/0004-6361/201832725}, 621, A42

\bibitem[\protect\citeauthoryear{Bardeen, Bond, Kaiser  \& Szalay}{Bardeen
  et~al.}{1986}]{Bardeen_StatisticsPeaksGaussian1986}
Bardeen J.~M.,  Bond J.~R.,  Kaiser N.,   Szalay A.~S.,  1986, \mn@doi [The
  Astrophysical Journal] {10.1086/164143}, 304, 15

\bibitem[\protect\citeauthoryear{Beattie, Federrath  \& Klessen}{Beattie
  et~al.}{2019}]{Beattie_RelationTrueObserved2019}
Beattie J.~R.,  Federrath C.,   Klessen R.~S.,  2019, \mn@doi [Monthly Notices
  of the Royal Astronomical Society] {10.1093/mnras/stz1416}, 487, 2070

\bibitem[\protect\citeauthoryear{Beattie, Mocz, Federrath  \& Klessen}{Beattie
  et~al.}{2021}]{Beattie_MultishockModelDensity2021}
Beattie J.~R.,  Mocz P.,  Federrath C.,   Klessen R.~S.,  2021, \mn@doi
  [Monthly Notices of the Royal Astronomical Society] {10.1093/mnras/stab1037},
  504, 4354

\bibitem[\protect\citeauthoryear{Bodenheimer}{Bodenheimer}{1995}]{Bodenheimer_AngularMomentumEvolution1995}
Bodenheimer P.,  1995, \mn@doi [Annual Review of Astronomy and Astrophysics]
  {10.1146/annurev.aa.33.090195.001215}, 33, 199

\bibitem[\protect\citeauthoryear{Bond \& Myers}{Bond \&
  Myers}{1996}]{Bond_PeakPatchPictureCosmic1996}
Bond J.~R.,  Myers S.~T.,  1996, \mn@doi [The Astrophysical Journal Supplement
  Series] {10.1086/192267}, 103, 1

\bibitem[\protect\citeauthoryear{Brunt, Federrath  \& Price}{Brunt
  et~al.}{2010}]{Brunt_MethodReconstructingVariance2010}
Brunt C.~M.,  Federrath C.,   Price D.~J.,  2010, \mn@doi [Monthly Notices of
  the Royal Astronomical Society] {10.1111/j.1365-2966.2009.16215.x}, 403, 1507

\bibitem[\protect\citeauthoryear{Burkhart}{Burkhart}{2018}]{Burkhart_StarFormationRate2018}
Burkhart B.,  2018, \mn@doi [The Astrophysical Journal]
  {10.3847/1538-4357/aad002}, 863, 118

\bibitem[\protect\citeauthoryear{{Chabrier}}{{Chabrier}}{2005}]{Chabrier_InitialMassFunction2005}
{Chabrier} G.,  2005, in {Corbelli} E.,  {Palla} F.,   {Zinnecker} H.,  eds,
  Astrophysics and Space Science Library Vol. 327, The Initial Mass Function 50
  Years Later. p.~41

\bibitem[\protect\citeauthoryear{Chandrasekhar}{Chandrasekhar}{1969}]{Chandrasekhar_EllipsoidalFiguresEquilibrium1969}
Chandrasekhar S.,  1969, Ellipsoidal figures of equilibrium.
\url {https://ui.adsabs.harvard.edu/abs/1969efe..book.....C}

\bibitem[\protect\citeauthoryear{Colman et~al.,}{Colman
  et~al.}{2022}]{Colman_SignatureLargeScale2022}
Colman T.,  et~al., 2022, \mn@doi [Monthly Notices of the Royal Astronomical
  Society] {10.1093/mnras/stac1543}, 514, 3670

\bibitem[\protect\citeauthoryear{Doroshkevich}{Doroshkevich}{1973}]{Doroshkevich_SpatialStructurePerturbations1973}
Doroshkevich A.~G.,  1973, \mn@doi [Astrophysics] {10.1007/BF01001625}, 6, 320

\bibitem[\protect\citeauthoryear{Eswaran \& Pope}{Eswaran \&
  Pope}{1988}]{Eswaran_ExaminationForcingDirect1988a}
Eswaran V.,  Pope S.,  1988, \mn@doi [Computers \& Fluids]
  {10.1016/0045-7930(88)90013-8}, 16, 257

\bibitem[\protect\citeauthoryear{Falgarone, Hily-Blant  \& Levrier}{Falgarone
  et~al.}{2004}]{Falgarone_StructureMolecularClouds2004}
Falgarone E.,  Hily-Blant P.,   Levrier F.,  2004, \mn@doi [Astrophysics and
  Space Science] {10.1023/B:ASTR.0000045004.70345.21}, 292, 89

\bibitem[\protect\citeauthoryear{Federrath}{Federrath}{2013}]{Federrath_UniversalitySupersonicTurbulence2013}
Federrath C.,  2013, \mn@doi [Monthly Notices of the Royal Astronomical
  Society] {10.1093/mnras/stt1644}, 436, 1245

\bibitem[\protect\citeauthoryear{Federrath}{Federrath}{2016}]{Federrath_UniversalityInterstellarFilaments2016}
Federrath C.,  2016, \mn@doi [Monthly Notices of the Royal Astronomical
  Society] {10.1093/mnras/stv2880}, 457, 375

\bibitem[\protect\citeauthoryear{Federrath \& Banerjee}{Federrath \&
  Banerjee}{2015}]{Federrath_DensityStructureStar2015}
Federrath C.,  Banerjee S.,  2015, \mn@doi [Monthly Notices of the Royal
  Astronomical Society] {10.1093/mnras/stv180}, 448, 3297

\bibitem[\protect\citeauthoryear{Federrath \& Klessen}{Federrath \&
  Klessen}{2013}]{Federrath_STARFORMATIONEFFICIENCY2013}
Federrath C.,  Klessen R.~S.,  2013, \mn@doi [The Astrophysical Journal]
  {10.1088/0004-637X/763/1/51}, 763, 51

\bibitem[\protect\citeauthoryear{Federrath, Klessen  \& Schmidt}{Federrath
  et~al.}{2009}]{Federrath_FRACTALDENSITYSTRUCTURE2009}
Federrath C.,  Klessen R.~S.,   Schmidt W.,  2009, \mn@doi [The Astrophysical
  Journal] {10.1088/0004-637X/692/1/364}, 692, 364

\bibitem[\protect\citeauthoryear{Federrath, Roman-Duval, Klessen, Schmidt  \&
  Mac~Low}{Federrath
  et~al.}{2010}]{Federrath_ComparingStatisticsInterstellar2010}
Federrath C.,  Roman-Duval J.,  Klessen R.~S.,  Schmidt W.,   Mac~Low M.-M.,
  2010, \mn@doi [Astronomy and Astrophysics] {10.1051/0004-6361/200912437},
  512, A81

\bibitem[\protect\citeauthoryear{Federrath, Klessen, Iapichino  \&
  Beattie}{Federrath et~al.}{2021}]{Federrath_SonicScaleInterstellar2021}
Federrath C.,  Klessen R.~S.,  Iapichino L.,   Beattie J.~R.,  2021, \mn@doi
  [Nature Astronomy] {10.1038/s41550-020-01282-z}, 5, 365

\bibitem[\protect\citeauthoryear{Ganguly, Walch, Seifried, Clarke  \&
  Weis}{Ganguly et~al.}{2023}]{Ganguly_UnravellingStructureMagnetized2023}
Ganguly S.,  Walch S.,  Seifried D.,  Clarke S.~D.,   Weis M.,  2023, \mn@doi
  [Monthly Notices of the Royal Astronomical Society] {10.1093/mnras/stad2054},
  525, 721

\bibitem[\protect\citeauthoryear{Ganguly, Walch, Clarke  \& Seifried}{Ganguly
  et~al.}{2024}]{Ganguly_SILCCZoomDynamicBalance2024}
Ganguly S.,  Walch S.,  Clarke S.~D.,   Seifried D.,  2024, \mn@doi [Monthly
  Notices of the Royal Astronomical Society] {10.1093/mnras/stae032}, 528, 3630

\bibitem[\protect\citeauthoryear{Hacar, Tafalla  \& Alves}{Hacar
  et~al.}{2017}]{Hacar_FibersNGC13332017a}
Hacar A.,  Tafalla M.,   Alves J.,  2017, \mn@doi [Astronomy \& Astrophysics]
  {10.1051/0004-6361/201630348}, 606, A123

\bibitem[\protect\citeauthoryear{Hacar, Clark, Heitsch, Kainulainen,
  Panopoulou, Seifried  \& Smith}{Hacar
  et~al.}{2022}]{Hacar_InitialConditionsStar2022}
Hacar A.,  Clark S.,  Heitsch F.,  Kainulainen J.,  Panopoulou G.,  Seifried
  D.,   Smith R.,  2022, Initial {Conditions} for {Star} {Formation}: {A}
  {Physical} {Description} of the {Filamentary} {ISM}, \url
  {http://arxiv.org/abs/2203.09562}

\bibitem[\protect\citeauthoryear{Hennebelle \& Chabrier}{Hennebelle \&
  Chabrier}{2008}]{Hennebelle_AnalyticalTheoryInitial2008}
Hennebelle P.,  Chabrier G.,  2008, \mn@doi [The Astrophysical Journal]
  {10.1086/589916}, 684, 395

\bibitem[\protect\citeauthoryear{Hennebelle \& Chabrier}{Hennebelle \&
  Chabrier}{2009}]{Hennebelle_ANALYTICALTHEORYINITIAL2009}
Hennebelle P.,  Chabrier G.,  2009, \mn@doi [The Astrophysical Journal]
  {10.1088/0004-637X/702/2/1428}, 702, 1428

\bibitem[\protect\citeauthoryear{Hennebelle \& Chabrier}{Hennebelle \&
  Chabrier}{2011}]{Hennebelle_ANALYTICALSTARFORMATION2011}
Hennebelle P.,  Chabrier G.,  2011, \mn@doi [The Astrophysical Journal]
  {10.1088/2041-8205/743/2/L29}, 743, L29

\bibitem[\protect\citeauthoryear{Hennebelle \& Chabrier}{Hennebelle \&
  Chabrier}{2013}]{Hennebelle_ANALYTICALTHEORYINITIAL2013}
Hennebelle P.,  Chabrier G.,  2013, \mn@doi [The Astrophysical Journal]
  {10.1088/0004-637X/770/2/150}, 770, 150

\bibitem[\protect\citeauthoryear{Hennebelle \& Falgarone}{Hennebelle \&
  Falgarone}{2012}]{Hennebelle_TurbulentMolecularClouds2012}
Hennebelle P.,  Falgarone E.,  2012, \mn@doi [The Astronomy and Astrophysics
  Review] {10.1007/s00159-012-0055-y}, 20, 55

\bibitem[\protect\citeauthoryear{Hopkins}{Hopkins}{2012}]{Hopkins_ExcursionsetModelStructure2012}
Hopkins P.~F.,  2012, \mn@doi [Monthly Notices of the Royal Astronomical
  Society] {10.1111/j.1365-2966.2012.20730.x}, 423, 2016

\bibitem[\protect\citeauthoryear{Hsieh, Arce, Mardones, Kong  \&
  Plunkett}{Hsieh et~al.}{2021}]{Hsieh_RotatingFilamentOrion2021}
Hsieh C.-H.,  Arce H.~G.,  Mardones D.,  Kong S.,   Plunkett A.,  2021, \mn@doi
  [The Astrophysical Journal] {10.3847/1538-4357/abd034}, 908, 92

\bibitem[\protect\citeauthoryear{Jaupart \& Chabrier}{Jaupart \&
  Chabrier}{2020}]{Jaupart_EvolutionDensityPDF2020}
Jaupart E.,  Chabrier G.,  2020, \mn@doi [The Astrophysical Journal]
  {10.3847/2041-8213/abbda8}, 903, L2

\bibitem[\protect\citeauthoryear{Jaupart \& Chabrier}{Jaupart \&
  Chabrier}{2022}]{Jaupart_StatisticalPropertiesCorrelation2022}
Jaupart E.,  Chabrier G.,  2022, \mn@doi [Astronomy \& Astrophysics]
  {10.1051/0004-6361/202141084}, 663, A113

\bibitem[\protect\citeauthoryear{Khullar, Federrath, Krumholz  \&
  Matzner}{Khullar et~al.}{2021}]{Khullar_DensityStructureSupersonic2021}
Khullar S.,  Federrath C.,  Krumholz M.~R.,   Matzner C.~D.,  2021, \mn@doi
  [Monthly Notices of the Royal Astronomical Society] {10.1093/mnras/stab1914},
  507, 4335

\bibitem[\protect\citeauthoryear{Krieger et~al.,}{Krieger
  et~al.}{2020}]{Krieger_TurbulentGasStructure2020}
Krieger N.,  et~al., 2020, \mn@doi [The Astrophysical Journal]
  {10.3847/1538-4357/aba903}, 899, 158

\bibitem[\protect\citeauthoryear{Kritsuk, Norman, Padoan  \& Wagner}{Kritsuk
  et~al.}{2007}]{Kritsuk_StatisticsSupersonicIsothermal2007}
Kritsuk A.~G.,  Norman M.~L.,  Padoan P.,   Wagner R.,  2007, \mn@doi [The
  Astrophysical Journal] {10.1086/519443}, 665, 416

\bibitem[\protect\citeauthoryear{Lacey \& Cole}{Lacey \&
  Cole}{1994}]{Lacey_MergerRatesHierarchical1994}
Lacey C.,  Cole S.,  1994, \mn@doi [Monthly Notices of the Royal Astronomical
  Society] {10.1093/mnras/271.3.676}, 271, 676

\bibitem[\protect\citeauthoryear{Larson}{Larson}{1981}]{Larson_TurbulenceStarFormation1981}
Larson R.~B.,  1981, \mn@doi [Monthly Notices of the Royal Astronomical
  Society] {10.1093/mnras/194.4.809}, 194, 809

\bibitem[\protect\citeauthoryear{Lomax, Whitworth  \& Cartwright}{Lomax
  et~al.}{2013}]{Lomax_IntrinsicShapesStarless2013}
Lomax O.,  Whitworth A.~P.,   Cartwright A.,  2013, \mn@doi [Monthly Notices of
  the Royal Astronomical Society] {10.1093/mnras/stt1764}, 436, 2680

\bibitem[\protect\citeauthoryear{Lombardi, Lada  \& Alves}{Lombardi
  et~al.}{2010}]{Lombardi_2MASSWideField2010}
Lombardi M.,  Lada C.~J.,   Alves J.,  2010, \mn@doi [Astronomy and
  Astrophysics] {10.1051/0004-6361/200912670}, 512, A67

\bibitem[\protect\citeauthoryear{Ludlow, Borzyszkowski  \& Porciani}{Ludlow
  et~al.}{2014}]{Ludlow_FormationCDMHaloes2014}
Ludlow A.~D.,  Borzyszkowski M.,   Porciani C.,  2014, \mn@doi [Monthly Notices
  of the Royal Astronomical Society] {10.1093/mnras/stu2021}, 445, 4110

\bibitem[\protect\citeauthoryear{Misugi, Inutsuka  \& Arzoumanian}{Misugi
  et~al.}{2022}]{Misugi_EvolutionAngularMomentum2022}
Misugi Y.,  Inutsuka S.-i.,   Arzoumanian D.,  2022, Evolution of the {Angular}
  {Momentum} of {Molecular} {Cloud} {Cores} {Formed} from {Filament}
  {Fragmentation}, \url {http://arxiv.org/abs/2212.02070}

\bibitem[\protect\citeauthoryear{Miville-Deschênes, Duc, Marleau, Cuillandre,
  Didelon, Gwyn  \& Karabal}{Miville-Deschênes
  et~al.}{2016}]{Miville-Deschenes_ProbingInterstellarTurbulence2016}
Miville-Deschênes M.-A.,  Duc P.-A.,  Marleau F.,  Cuillandre J.-C.,  Didelon
  P.,  Gwyn S.,   Karabal E.,  2016, \mn@doi [Astronomy \& Astrophysics]
  {10.1051/0004-6361/201628503}, 593, A4

\bibitem[\protect\citeauthoryear{Molina, Glover, Federrath  \& Klessen}{Molina
  et~al.}{2012}]{Molina_DensityVarianceMachNumber2012}
Molina F.~Z.,  Glover S. C.~O.,  Federrath C.,   Klessen R.~S.,  2012, \mn@doi
  [Monthly Notices of the Royal Astronomical Society]
  {10.1111/j.1365-2966.2012.21075.x}, 423, 2680

\bibitem[\protect\citeauthoryear{Passot \& Vázquez-Semadeni}{Passot \&
  Vázquez-Semadeni}{1998}]{Passot_DensityProbabilityDistribution1998}
Passot T.,  Vázquez-Semadeni E.,  1998, \mn@doi [Physical Review E]
  {10.1103/PhysRevE.58.4501}, 58, 4501

\bibitem[\protect\citeauthoryear{Pineda et~al.,}{Pineda
  et~al.}{2022}]{Pineda_BubblesFilamentsCores2022}
Pineda J.~E.,  et~al., 2022, From {Bubbles} and {Filaments} to {Cores} and
  {Disks}: {Gas} {Gathering} and {Growth} of {Structure} {Leading} to the
  {Formation} of {Stellar} {Systems}, \url {http://arxiv.org/abs/2205.03935}

\bibitem[\protect\citeauthoryear{Pineda et~al.,}{Pineda
  et~al.}{2024}]{Pineda_ProbingPhysicsStarFormation2024}
Pineda J.~E.,  et~al., 2024, \mn@doi [Astronomy \& Astrophysics]
  {10.1051/0004-6361/202451208}, 690, L5

\bibitem[\protect\citeauthoryear{Press \& Schechter}{Press \&
  Schechter}{1974}]{Press_FormationGalaxiesClusters1974}
Press W.~H.,  Schechter P.,  1974, \mn@doi [The Astrophysical Journal]
  {10.1086/152650}, 187, 425

\bibitem[\protect\citeauthoryear{Punanova, Caselli, Pineda, Pon, Tafalla, Hacar
   \& Bizzocchi}{Punanova et~al.}{2018}]{Punanova_KinematicsDenseGas2018}
Punanova A.,  Caselli P.,  Pineda J.~E.,  Pon A.,  Tafalla M.,  Hacar A.,
  Bizzocchi L.,  2018, \mn@doi [Astronomy \& Astrophysics]
  {10.1051/0004-6361/201731159}, 617, A27

\bibitem[\protect\citeauthoryear{Rabatin \& Collins}{Rabatin \&
  Collins}{2023}]{Rabatin_DensityVelocityCorrelations2023}
Rabatin B.,  Collins D.~C.,  2023, Density and {Velocity} {Correlations} in
  {Isothermal} {Supersonic} {Turbulence}, \url
  {http://arxiv.org/abs/2307.04876}

\bibitem[\protect\citeauthoryear{Robitaille, Motte, Schneider, Elia  \&
  Bontemps}{Robitaille et~al.}{2019a}]{Robitaille_ExposingPluralNature2019a}
Robitaille J.-F.,  Motte F.,  Schneider N.,  Elia D.,   Bontemps S.,  2019a,
  \mn@doi [Astronomy \& Astrophysics] {10.1051/0004-6361/201935545}, 628, A33

\bibitem[\protect\citeauthoryear{Robitaille, Motte, Schneider, Elia  \&
  Bontemps}{Robitaille et~al.}{2019b}]{Robitaille_ExposingPluralNature2019}
Robitaille J.-F.,  Motte F.,  Schneider N.,  Elia D.,   Bontemps S.,  2019b,
  \mn@doi [Astronomy \& Astrophysics] {10.1051/0004-6361/201935545}, 628, A33

\bibitem[\protect\citeauthoryear{Schmidt, Hillebrandt  \& Niemeyer}{Schmidt
  et~al.}{2006}]{Schmidt_NumericalDissipationBottleneck2006a}
Schmidt W.,  Hillebrandt W.,   Niemeyer J.~C.,  2006, \mn@doi [Computers \&
  Fluids] {10.1016/j.compfluid.2005.03.002}, 35, 353

\bibitem[\protect\citeauthoryear{Schmidt, Federrath, Hupp, Kern  \&
  Niemeyer}{Schmidt
  et~al.}{2009}]{Schmidt_NumericalSimulationsCompressively2009}
Schmidt W.,  Federrath C.,  Hupp M.,  Kern S.,   Niemeyer J.~C.,  2009, \mn@doi
  [Astronomy \& Astrophysics] {10.1051/0004-6361:200809967}, 494, 127

\bibitem[\protect\citeauthoryear{Sheth \& Tormen}{Sheth \&
  Tormen}{2002}]{Sheth_ExcursionSetModel2002}
Sheth R.~K.,  Tormen G.,  2002, \mn@doi [Monthly Notices of the Royal
  Astronomical Society] {10.1046/j.1365-8711.2002.04950.x}, 329, 61

\bibitem[\protect\citeauthoryear{Sheth, Mo  \& Tormen}{Sheth
  et~al.}{2001}]{Sheth_EllipsoidalCollapseImproved2001}
Sheth R.~K.,  Mo H.~J.,   Tormen G.,  2001, \mn@doi [Monthly Notices of the
  Royal Astronomical Society] {10.1046/j.1365-8711.2001.04006.x}, 323, 1

\bibitem[\protect\citeauthoryear{Tanaka, Nagai, Kamegai, Iino  \& Sakai}{Tanaka
  et~al.}{2020}]{Tanaka_HCNHNC132020}
Tanaka K.,  Nagai M.,  Kamegai K.,  Iino T.,   Sakai T.,  2020, \mn@doi [The
  Astrophysical Journal] {10.3847/1538-4357/abbcca}, 903, 111

\bibitem[\protect\citeauthoryear{Tatematsu, Ohashi, Sanhueza, Nguyen~Luong,
  Umemoto  \& Mizuno}{Tatematsu
  et~al.}{2016}]{Tatematsu_AngularMomentumCores2016}
Tatematsu K.,  Ohashi S.,  Sanhueza P.,  Nguyen~Luong Q.,  Umemoto T.,   Mizuno
  N.,  2016, \mn@doi [Publications of the Astronomical Society of Japan]
  {10.1093/pasj/psw002}, 68, 24

\bibitem[\protect\citeauthoryear{Teyssier}{Teyssier}{2002}]{Teyssier_CosmologicalHydrodynamicsAdaptive2002}
Teyssier R.,  2002, \mn@doi [Astronomy \& Astrophysics]
  {10.1051/0004-6361:20011817}, 385, 337

\bibitem[\protect\citeauthoryear{Vazquez-Semadeni}{Vazquez-Semadeni}{1994}]{Vazquez-Semadeni_HierarchicalStructureNearly1994}
Vazquez-Semadeni E.,  1994, \mn@doi [The Astrophysical Journal]
  {10.1086/173847}, 423, 681

\bibitem[\protect\citeauthoryear{White \& Silk}{White \&
  Silk}{1979}]{White_GrowthAsphericalStructure1979}
White S. D.~M.,  Silk J.,  1979, \mn@doi [The Astrophysical Journal]
  {10.1086/157156}, 231, 1

\makeatother
\end{thebibliography}




\appendix

\section{Correlation function and second order momentum}
\label{App_Correlation_function}

We consider a logdensity power spectrum  scaling as $P (k)\propto k^{-4}$ for a turbulent medium as suggested by simulations. The correlation function is:
\begin{align}
	g^R_s(q)&=\int_0^{+\infty} P(k) W_R(k) \sinc(kq)2\pi k^2 {\rm d}k, \\
    &=\int_{a=\frac{2\pi}{L_{\rm i}}}^{b=\frac{2\pi}{R}}Ck^{-2}\sinc(kq){\rm d}k.
\end{align}
In this Appendix, we note the spatial lag \(q\) instead of \(\Delta q\) to simplify the notations.
To determine the constant, we assume that the dispersion of the turbulence at the local scale \(\sigma_s^2=\beta^2\mathcal{M}^2\) (see e.g. \cite{Molina_DensityVarianceMachNumber2012}) is the one reached at the dissipation scale of the turbulence cascade \(R_{\rm dis}=\frac{2\pi}{b_{\rm dis}}\). Thus:
\begin{align}
	\sigma_s^2 &=C\int_{a=\frac{2\pi}{L_{\rm i}}}^{b=\frac{2\pi}{R_{\rm dis}}}k^{-2}{\rm d}k, \\
	C &= \frac{ab_{\rm dis}}{b_{\rm dis}-a}\sigma_s^2\simeq a\sigma_s^2,
	\label{C_sigma}
\end{align}
when the dissipation scale is much smaller than the injection scale. It is likely to be the case because the natural dissipation scale for turbulence fluctuations is of the order of the sonic length, more than two orders of magnitude smaller than a typical injection scale of $\sim$10 pc. For \(b<b_{\rm dis}\), the correlation function is given by
\begin{multline}
    \label{Eq_corr_func_app}
	g^R_s(q)=\frac{\sigma_s^2}{2}\left(\sinc (a q)-\frac{a}{b}\sinc (b q)+\cos (a q)-\frac{a}{b}\cos (b q)\right. \\
	+qa \text{Si}(a q)-qa\text{Si}(b q)\bigg)
\end{multline}
where \(\text{Si}\) is the hyperbolic sinus function.

The dispersion of the derivatives of the field \(s\) at the maximum are related to the correlation function by \(\mathbb{E}((\eta_i^R)^2) = -(g_s^R)''(0)\) and \(\mathbb{E}((\mathbb{I}_{ii}^R)^2) = (g_s^R)^{(4)}(0)\). They can be simply computed by the asymptotic expansion around 0.
\begin{multline}
	g^R_s(q)=\frac{C}{360}(b^3-a^3) q^4-\frac{C}{6} (b-a)q^2+C(\frac{1}{a}-\frac{1}{b}).
\end{multline}

Finally:
\begin{align}
	\mathbb{E}((\eta_i^R)^2) &= \frac{C}{3} (b-a) \simeq\frac{\sigma_s^2}{3}(ba-a^2), \\
	\mathbb{E}((\mathbb{I}_{ii}^R)^2) &= \frac{C}{15}(b^3-a^3) \simeq\frac{\sigma_s^2}{15}(ab^3-a^4).
\end{align}

To avoid extra numerical coefficients in the derivation of the PDF of the eigenvalues, we define \(\sigma_\eta^2(R)=3\mathbb{E}((\eta_i^R)^2)\) and \(\sigma_\mathbb{I}^2(R)=5\mathbb{E}((\mathbb{I}_{ii}^R)^2)\) which are given in equations \ref{Sigma_1} and \ref{Sigma_2}.

\section{Convergence Study and impact of the extraction parameters}
\label{App_Convergence}

In this appendix, we test the convergence of the numerical verification of our model presented in \S\ref{Sec_Numerical_verification}. We first vary the numerical resolution and then the ratio \(\rho_{\rm max}/\rho_{\rm min}\).

\begin{figure}
    \centering
    \includegraphics[width=\hsize]{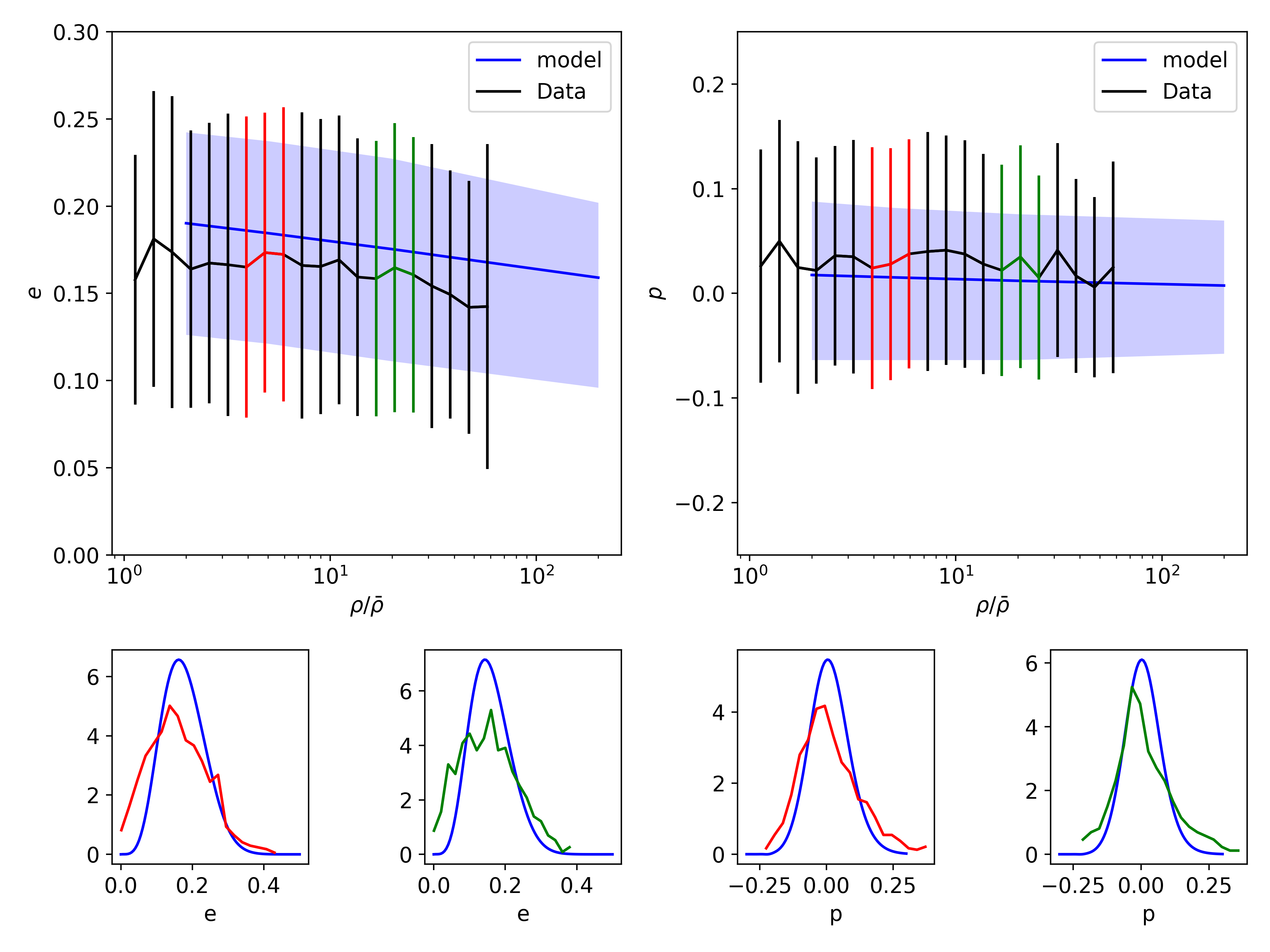}
       \caption{Same as Fig. \ref{Fig_numerical_verification} but in a 512\(^3\) run and each semi-axes is resolved by at least 4 cells.
               }
          \label{Fig_convergence_resolution}
\end{figure}

In Fig. \ref{Fig_convergence_resolution}, we plot the mean and standard deviation of the ellipticity and prolateness measured in a 512\(^3\) simulation characterised by the same properties as in \S\ref{Sec_Numerical_verification}. Here, however, in order to extract structures of the same physical size as those extracted in the 1024\(^3\), we impose that each axis must be resolved by at least 4 cells. Compared to Fig. \ref{Fig_numerical_verification}, the ellipticity and the descreening decrease by about 10\% when the numerical resolution is increased. We can conclude that the convergence is reasonably well achieved. 

\begin{figure}
    \centering
    \includegraphics[width=\hsize]{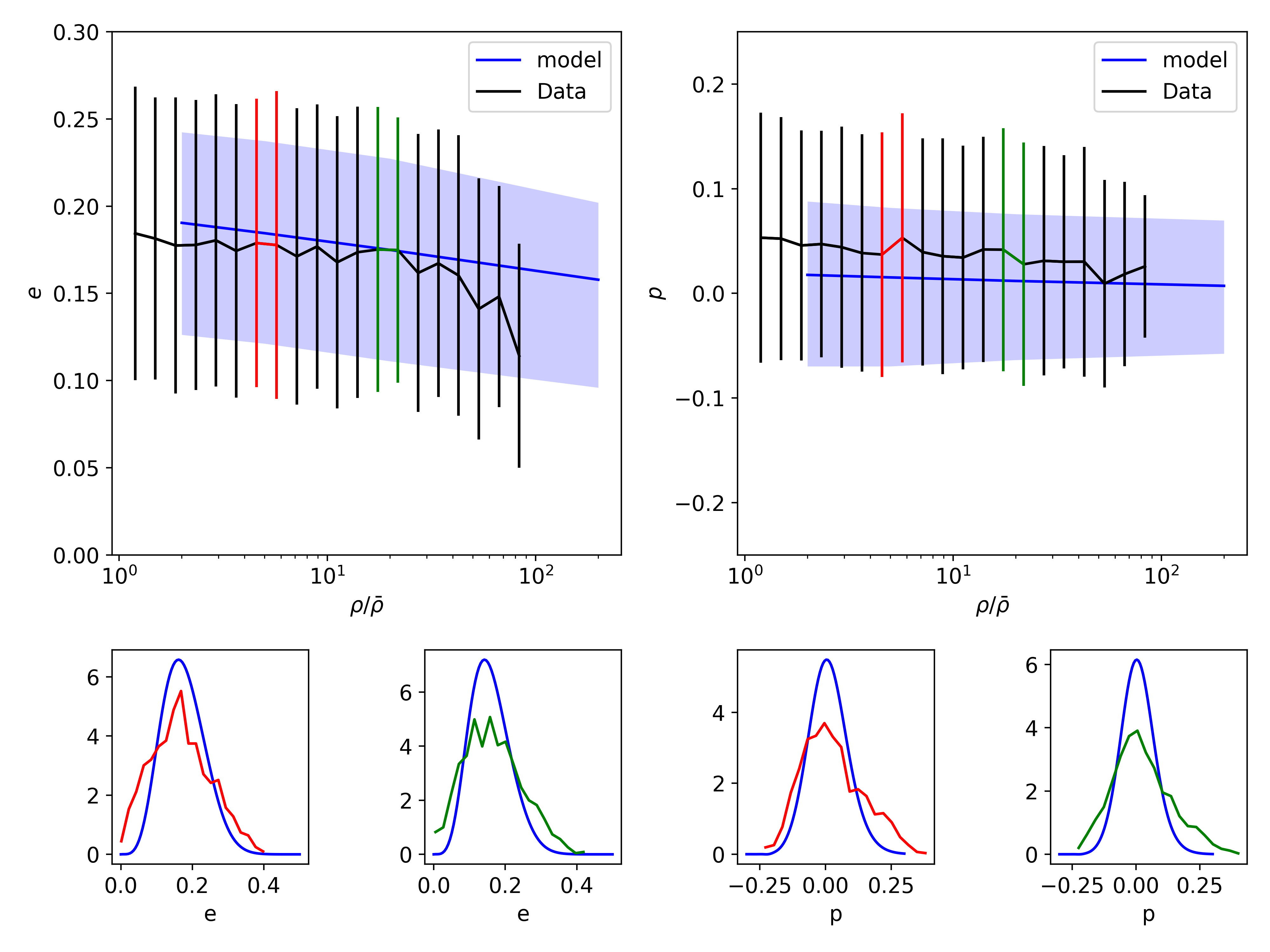}
       \caption{Same as Figure \ref{Fig_numerical_verification} but with \(\rho_{\rm max}/\rho_{\rm min}=2\)
               }
          \label{Fig_convergence_drho}
\end{figure}

In Fig. \ref{Fig_convergence_drho}, we plot the mean and standard deviation of the ellipticity and prolateness measured in a 1024\(^3\) imposing that \(\rho_{\rm max}/\rho_{\rm min}=2\) and each axis must be resolved by at least 8 cells. Although there are some variations in the statistics of the extracted structures compared to Fig. \ref{Fig_convergence_resolution}, the model still reproduces the measurements well. 

\section{Impact of the rotation of prestellar clumps}
\label{Rotation}
In addition to the turbulent support, the prestellar clumps are supported by their rotational energy. From the filament scale to the core scale, observations are roughly in agreement with a dependence of the specific angular momentum with the radius of the structure between \(j(r)\propto r^{1.5}\) and \(j(r)\propto r^{2.2}\) \citep{Bodenheimer_AngularMomentumEvolution1995, Tatematsu_AngularMomentumCores2016,Punanova_KinematicsDenseGas2018, Hsieh_RotatingFilamentOrion2021, Misugi_EvolutionAngularMomentum2022}. Here, we will use the relation given by \cite{Tatematsu_AngularMomentumCores2016}:
\begin{equation}
	\label{spectific_angular_momentum}
	j=10^{21.4}\left(\frac{R}{0.1 pc}\right)^\alpha \text{cm}^2\cdot\text{s}^{-1},
\end{equation}
with \(\alpha=1.65\). We assume that this relation is also valid at the very initial stage of the clump formation. The Virial equilibrium considering the rotation energy can be written as: 
\begin{equation}
	\label{Rotation_Virial}
	\frac{2}{5}\pi G\bar{\rho}\e^{s_R^{\rm c}}a_2^2 I-\frac{a_2^2\Omega^2}{5}\left(1+\frac{a_1^2}{a_2^2}+\frac{a_1^2}{a_3^2}\right)=<V_{\rm RMS}^2>+3C_{\rm s}^2,
\end{equation}
where \(\Omega\) is the angular velocity of the clump. With equation \ref{spectific_angular_momentum}, we have
\begin{equation}
	a_2\Omega=0.03 \left(\frac{R}{0.1\text{pc}}\right)^{\alpha-1}\text{km/s}.
\end{equation}
This rotational speed is approximately one order of magnitude smaller than the velocity dispersion arising from transonic or supersonic turbulence. Moreover, from \ref{Rotation_Virial}, the rotational term involved in the equation of the barrier is proportional to \(j/R^2\propto R^{\alpha-2}\). As \(\alpha\simeq 2\) in most of the observations, the rotational term is independent of the formation scale of the prestellar clumps. Thus, it increases uniformly the barrier and thus barely affects the shape of the CMF.

\section{CMF for other injections lengths}
\label{Other_injection_lengths}

\begin{figure}
	\centering
	\includegraphics[width=\hsize]{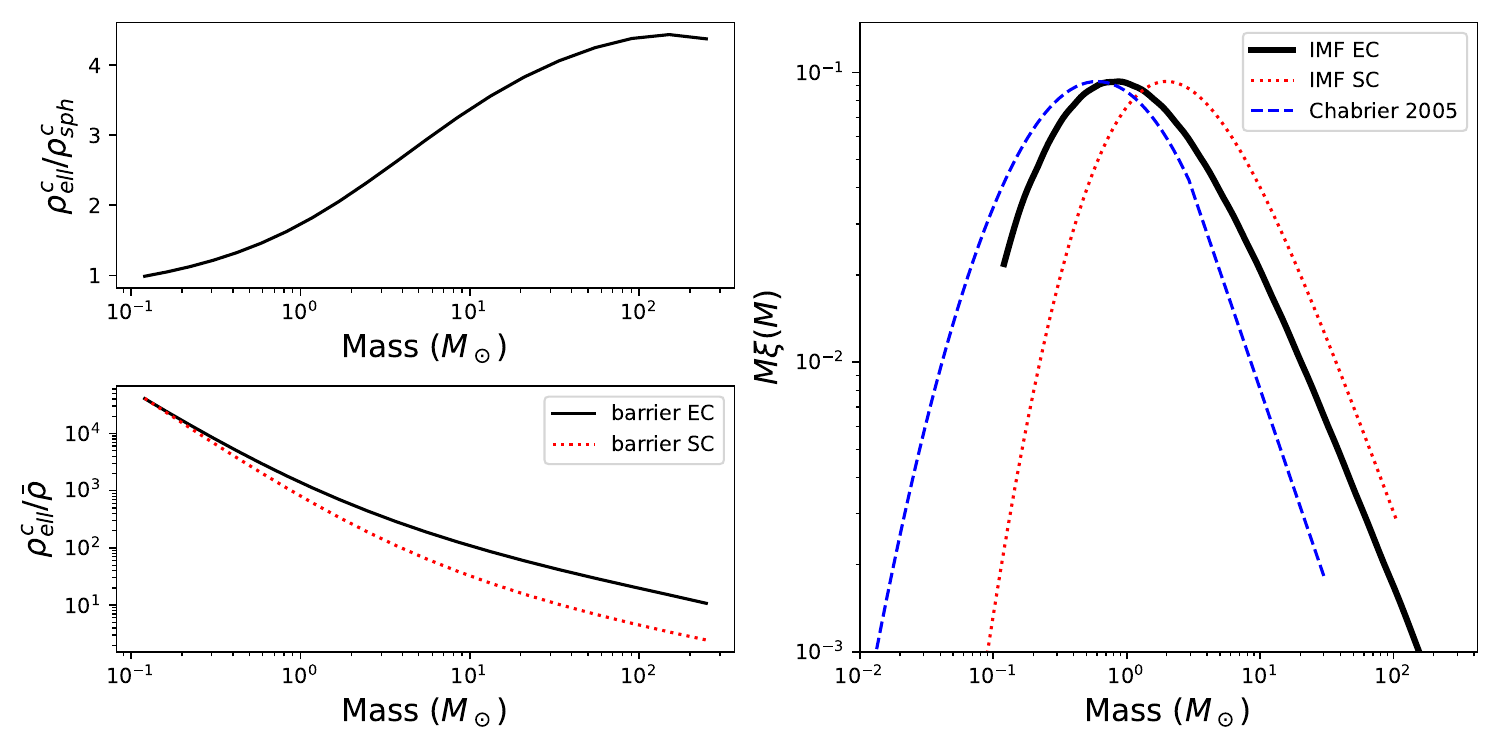}
	   \caption{Same as Fig. \ref{fig_CMF} for a turbulent injection scale \(L_{\rm i}=20\)pc.}
		\label{fig_CMF_20}
\end{figure}

To cover the whole range of masses initially triggered by turbulence induced fluctuations, different injection scales should be considered. In Fig. \ref{fig_CMF_20}, we plot the CMF of structures formed by turbulence in the time independent case and injected at \(L_{\rm i}=20\) pc. We confirm that, whatever the injection scale, the CMF has a power law slightly shallower than the Salpeter one, due to the more pronounced deformation of the less dense objects.

\bsp	
\label{lastpage}
\end{document}